\documentclass{aa}
\usepackage{multirow}

\usepackage{multirow}
\usepackage{amsmath}
\usepackage[colorlinks,linkcolor=blue,anchorcolor=blue,citecolor=blue]{hyperref}
\usepackage{graphicx,subfig} 
\usepackage{lscape}
\usepackage{longtable}
\usepackage{txfonts}
\usepackage{natbib}
\usepackage{color}
\usepackage{booktabs}

\newcommand{\rosi}{\textit{eROSITA}\xspace}
\newcommand{\xmm}{\textit{XMM-Newton}\xspace}
\newcommand{\chandra}{\textit{Chandra}\xspace}

\usepackage{natbib,twoopt}
\usepackage{xcolor}

\makeatletter
\newcommandtwoopt{\citeads}[3][][]{\href{http://adsabs.harvard.edu/abs/#3}%
    {\def\hyper@linkstart##1##2{}%
     \let\hyper@linkend\@empty\citealp[#1][#2]{#3}}}
  \newcommandtwoopt{\citepads}[3][][]{\href{http://adsabs.harvard.edu/abs/#3}%
    {\def\hyper@linkstart##1##2{}%
     \let\hyper@linkend\@empty\citep[#1][#2]{#3}}}
  \newcommandtwoopt{\citetads}[3][][]{\href{http://adsabs.harvard.edu/abs/#3}%
    {\def\hyper@linkstart##1##2{}%
     \let\hyper@linkend\@empty\citet[#1][#2]{#3}}}
  \newcommandtwoopt{\citeyearads}[3][][]%
    {\href{http://adsabs.harvard.edu/abs/#3}
    {\def\hyper@linkstart##1##2{}%
     \let\hyper@linkend\@empty\citeyear[#1][#2]{#3}}}
\makeatother
\usepackage{graphicx}
\usepackage{multirow}
\usepackage{siunitx}  
\usepackage{txfonts}

\begin{document}

\title{The eROSITA Final Equatorial-Depth Survey (eFEDS):}
\subtitle{Galaxy Clusters and Groups in Disguise }

\author{
Esra~Bulbul\inst{1}\thanks{e-mail: \href{mailto:ebulbul@mpe.mpg.de}{\tt ebulbul@mpe.mpg.de}},
Ang~Liu \inst{1},
Thomas~Pasini \inst{2},
Johan~Comparat \inst{1},
Duy~N.~Hoang \inst{2},
Matthias~Klein \inst{3},
Vittorio~Ghirardini \inst{1},
Mara~Salvato \inst{1},
Andrea~Merloni\inst{1}, 
Riccardo~Seppi\inst{1},
Julien~Wolf\inst{1,4},
Scott~F.~Anderson \inst{5},
Y.~Emre~Bahar\inst{1},
Marcella~Brusa\inst{6,7},
Marcus~Br\"{u}ggen\inst{2},
Johannes~Buchner\inst{1},
Tom~Dwelly \inst{1}, 
Hector~Ibarra-Medel\inst{8},
Jacob~Ider~Chitham \inst{1},
Teng~Liu \inst{1},
Kirpal~Nandra \inst{1}, 
Miriam~E.~Ramos-Ceja\inst{1},
Jeremy~S.~Sanders\inst{1},
Yue~Shen\inst{9}
}

\institute{
Max Planck Institute for Extraterrestrial Physics, Giessenbachstrasse 1, 85748 Garching, Germany 
\and
University of Hamburg, Hamburger Sternwarte, Gojenbergsweg 112, 21029 Hamburg, Germany
\and
Universitaets-Sternwarte Muenchen, Fakultaet fuer Physik, LMU Munich, Scheinerstr. 1, 81679 Munich, Germany 
\and 
Exzellenzcluster ORIGINS, Boltzmannstr. 2, D-85748 Garching, Germany
\and 
University of Washington Department of Astronomy Box 351580, Seattle, WA 98195, USA
\and 
Dipartimento di Fisica e Astronomia "Augusto Righi", Universit\'a di Bologna,  via Gobetti 93/2,  40129 Bologna, Italy 
\and INAF - Osservatorio di Astrofisica e Scienza dello Spazio di Bologna, via Gobetti 93/3,  40129 Bologna, Italy 
\and
Department of Astronomy, University of Illinois at Urbana-Champaign, Urbana, IL 61801, USA
\and 
University of Illinois at Urbana-Champaign Department of Astronomy, Urbana, IL 61801, USA
}

\titlerunning{eFEDS galaxy clusters and groups in disguise}
\authorrunning{Bulbul et al.}

\abstract
 {}
{The eROSITA Final Equatorial-Depth Survey (eFEDS), executed during the performance verification phase of the Spectrum-Roentgen-Gamma (SRG)/eROSITA telescope, was completed in November 2019. One of the science goals of this survey is to demonstrate the ability of \rosi\ to detect samples of clusters and groups at the final depth of the \rosi\ all-sky survey. }
{Because of the sizeable ($\approx 26^{\prime\prime}$ HEW FOV average) point-spread function of \rosi, high-redshift clusters of galaxies or compact nearby groups hosting bright active galactic nuclei (AGN) can be misclassified as point sources by the source detection algorithms. A total of 346 galaxy clusters and groups in the redshift range of $0.1<z<1.3$ were identified based on their red sequence in the eFEDS point source catalog. }
{We examine the multiwavelength properties of these clusters and groups to understand the potential biases in our selection process and the completeness of the extent-selected sample. We find that the majority of the clusters and groups in the point source sample are indeed underluminous and compact compared to the extent-selected sample. Their faint X-ray emission, well below the flux limit of the extent-selected eFEDS clusters, and their compact X-ray emission are likely to be the main reason for this misclassification. In the sample, we confirm that 10\% of the sources host AGN in their brightest cluster galaxies (BCGs) through optical spectroscopy and visual inspection. By studying their X-ray, optical, infrared, and radio properties, we establish a method for identifying clusters and groups that host AGN in their BCGs. We successfully test this method on the current point source catalog through the Sloan Digital Sky Survey optical spectroscopy and find eight low-mass clusters and groups with active radio-loud AGN that are particularly bright in the infrared. They include eFEDS~J091437.8+024558, eFEDS~J083520.1+012516, and eFEDS~J092227.1+043339 at redshifts 0.3-0.4.}
{This study helps us to characterize and understand our selection process and assess the completeness of the \rosi\ extent-selected samples. The method we developed will be used to identify high-redshift clusters, AGN-dominated groups, and low-mass clusters that are misclassified in the future \rosi\ all-sky survey point source catalogs. }

\maketitle

\section{Introduction}

Galaxy clusters are located at the peaks in the cosmic density field and offer an independent and powerful probe of the growth of structure. Their overall abundance in the Universe is strongly dependent on the underlying cosmology. Extending the samples of known clusters of galaxies through ground-based and space-based observatories therefore is a way forward to place tight constraints on the cosmological parameters based on cluster counts \citep{Kravtsov2012}. Cluster surveys performed with ground-based optical telescopes relying on finding overdensities of red galaxies in the sky suffer from projection effects \citep{Farahi2016, Zu2017, Busch2017}. Owing to the redshift-independent signal, cluster surveys based on the Sunyaev-Zeldovich (SZ) effect \citep{sz1972} performed with the {\sl Planck} Space Telescope, the South Pole Telescope (SPT), and the Atacama Cosmology Telescope (ACT) provide an advantage over detecting clusters out to high redshifts \citep{Planck2014, Bleem2015, Hilton2021}. However, their ability to detect nearby clusters of galaxies is limited by their beam size \citep{Hincks2010, Planck2011}.

X-ray surveys are less sensitive to projection effects than optical surveys, and they have a better spatial resolution than the current SZ surveys. Therefore, they offer an efficient way for detecting clusters of galaxies. Despite these advantages, their selection function is crucial for further science exploitation \citep{Clerc2018}. Derivation of the selection function for extended X-ray surveys relies on the detection of the X-ray sources and their classification as extended objects. The first X-ray all-sky survey (RASS) was performed with the ROSAT telescope and provided the extensive cluster catalogs obtained in the soft X-ray band \citep{Ebeling2001, Boehringer2000, Boehringer2004}. Since the early 2010s, smaller but deeper X-ray surveys performed with \xmm\ provided samples of clusters down to a flux limit of $10^{-15}$~erg~s$^{-1}$~cm$^{-2}$ and allowed determining cosmological parameters through accurate modeling of the X-ray selection function \citep{Pacaud2016, Adami2018}. With the new availability of the extended ROentgen Survey with an Imaging Telescope Array (\rosi) All-Sky Survey \citep[eRASS,][]{predehl2021} in the full 0.2--10~keV X-ray band, we are on the verge of detecting 100,000 clusters and groups of galaxies down to the mass limit of $10^{13}\ M_{sun}$, and we will be able to constrain cosmological parameters with a percent-level precision through cluster counts \citep{Merloni2012}.

Clusters of galaxies appear as extended sources in the X-ray sky mainly due to Bremsstrahlung emission from their diffuse intracluster medium (ICM). The source characterization strategy of the X-ray cluster surveys of \rosi\ relies on the detection of extended sources based on their extent and detection likelihoods \citep{Brunner2021, Liu2021}. Identification of the counterparts to these sources is then performed with multiple optical surveys and strategies \citep[e.g.,][]{Klein2021, Salvato2021}. Because of the sizable point spread functions (PSF) of X-ray telescopes, clusters of galaxies that host bright central active galactic nuclei (AGN) or peaked cool-core emission at high redshifts, or faint compact extended emission, can be missed by the selection cuts and misclassified as point sources \citep{Green2017, Biffi2018, Clerc2018, Ghirardini2021c}. We note that, in the context of this work, we define groups as halos with masses $M_{500}<10^{14}M_{sun}$. For instance, it has been shown that very extended groups that have a central AGN may be missed by the extent selection in the ROSAT catalog because their surface brightness profiles are shallower than the beta model that is used as a template by the source detection algorithm \citep{Xu2018}. It is vital to understand the reasons for the misclassification and to determine the total number of these cases to understand the uncertainties in the selection function and the completeness of the \rosi\ survey. 

Beyond testing the selection function, clusters hosting bright AGN in their cores are also interesting targets for astrophysical investigations and searches for dark matter. AGN outflows and jets play an important role in the formation and evolution of galaxies, in establishing the multiphase environment from group scales to clusters, and in regulating star formation and cooling flows in cluster cores \citep[e.g.,][]{Rafferty2006,Gaspari2012,Gaspari2017}. In the local Universe, this feedback power is sufficient to offset the strong radiative cooling in the cores of clusters \citep{mcnamara2007}, although the details of this feedback process are still a matter of much debate \citep[see][for a review]{Fabian2012}. It is vital to study feedback in groups and clusters with bright central AGN to understand how feedback influences the thermal history of the matter in these dense environments. Moreover, the mechanical output of radio feedback can be measured from the properties of the jet-inflated radio-plasma-filled cavities in the ICM \citep[e.g.,][]{Churazov2000,birzan2004}, a joint analysis combining X-ray and radio data of these clusters is therefore important to complete the picture of AGN feedback.
Clusters of galaxies that host bright central AGN can also be used to constrain the parameter space of axion-like particles, an ultralight theoretically well-motivated cold dark matter candidate \citep[e.g.,][]{Burrage2009,Horns2012}. Axion-like particles and photons interconvert in the presence of magnetic fields, leaving imprints on the X-ray spectra of sources. AGN embedded in large dark matter concentrations, such as galaxy clusters, provide an ideal environment with a magnetic field for this conversion to occur \citep{Conlon2017, Berg2017, Reynolds2020}. 

One famous example in which a cluster is characterized as a point source in the ROSAT all-sky survey is the Phoenix cluster, which was later detected as an extended source through its significant SZ decrement in the millimeter-wave SPT data \citep{McDonald2012}. Previously known clusters that host a bright AGN  include A1835 \citep[$z=0.25$;][]{McNamara2006}, NGC~5813 \citep[$z=0.0065$;][]{Randall2015}, MACS~J1931.8-2634 \citep[$z=0.35$;][]{Ehlert2011}, and 3C~186 \citep[$z=1.07$][]{Siemiginowska2005}. In recent years, great effort has been made to identify more clusters that are  concealed by the hosting bright AGN and star-forming galaxies in the ROSAT all-sky survey point source catalog \citep{Green2017, Donahue2020, Somboonpanyakul2021b}. In these searches, a few remarkable clusters have been discovered, such as CHIPS~1356-3421 ($z=0.22$) and CHIPS~1911+4455 ($z=0.48$) \citep{Somboonpanyakul2018, Somboonpanyakul2021a}. Another 14 clusters hosting X-ray bright brightest cluster galaxies (BCGs) were identified in \citet{Donahue2020}. The next logical step is to identify more of these extreme systems to extend feedback studies to groups with lower mass and clusters at higher redshifts, where we expect to see significant AGN activity, by pushing the flux limit down with the \rosi\ all-sky survey.

\begin{figure*}[ht!]
\begin{center}
\includegraphics[width=0.99\textwidth, trim=60 10 35 50, clip]{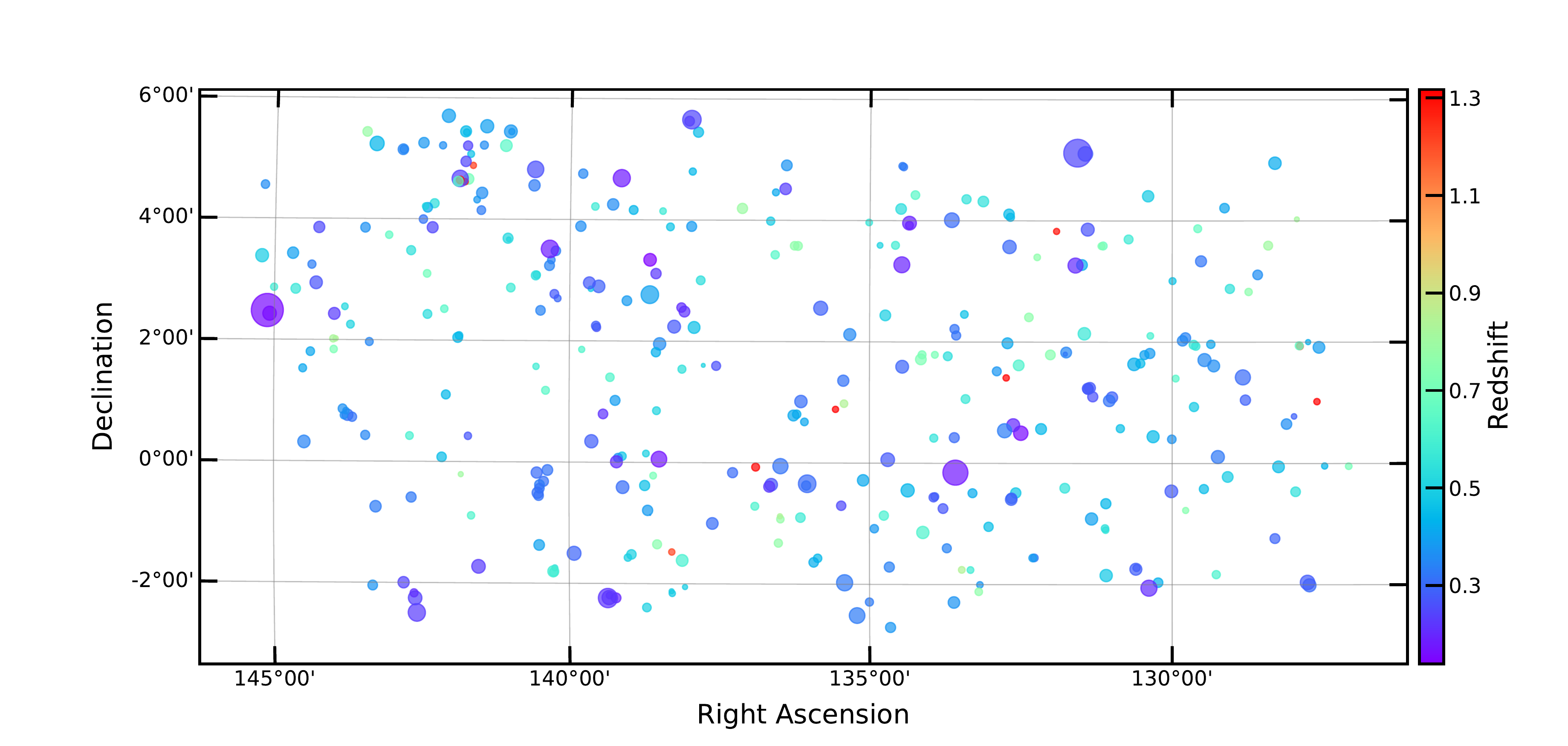}
\caption{Distribution of the 346 cluster candidates in eFEDS point source catalog. The color code represents the redshift of the cluster, provided by the MCMF \citep[  ][]{Klein2018,Klein2019}. The radius of the circle corresponds to $3R_{500}$ of each cluster.}
\label{fig:cat}
\end{center}
\end{figure*}

The \rosi\ Final Equatorial-Depth Survey (eFEDS), completed during the performance verification phase of \rosi,\ is designed to demonstrate the survey science capabilities of the final all-sky survey \citep{predehl2021}. The eFEDS field covers a 140 deg$^2$ area in the equatorial region of eRASS (126$^{\circ}$ < R.A. < 146$^{\circ}$ and -3$^{\circ}$ < Dec. < +6$^{\circ}$) and has a survey depth with a vignetted exposure time of 1.2~ks in the 0.5--2.0~keV energy band. The full X-ray source catalog including point and extended sources was presented in \citet{Brunner2021}. Using the extent and detection likelihood selections, \citet{Liu2021} compiled a sample of clusters of galaxies in the eFEDS field. The characterization and follow-up of the extended sources in this main catalog were presented in  \citet{Liu2021}, \citet{Ghirardini2021b}, and \citet{Klein2021}. A point source catalog was also compiled based on the X-ray detection properties \citep{Liut2021}, and the optical follow-up, characterization, and classification of these sources were presented in S21 and Nandra et al. (2021, in prep), respectively. \citet[][S21 hereafter]{Salvato2021} found that 346 of the 27369 point-like sources are secure clusters of galaxies with probabilities higher than 80\%. This result is based on concentrations of red sequence galaxies around the X-ray position and on the comparison of photometric redshifts computed with two independent methods under different assumptions.

The primary goal of this paper is to characterize the properties of these candidate clusters and groups in the eFEDS point source sample, identified by S21, based on the multiwavelength data in X-ray, optical, and radio bands to understand the selection and completeness of the \rosi\ cluster surveys. Another goal of this study is to identify clusters of galaxies or groups that have a central AGN, which are ideal targets for studying the AGN feedback cycle and constraining the parameter space for the axion-like particles. The eFEDS field has also been observed with a broad array of multiwavelength survey instruments through the LOw Frequency Array (LOFAR) in radio, optical photometric, and spectroscopic follow-up with the Dark Energy Camera Legacy Survey (DECaLS), Hyper Supreme Camera (HSC), the Sloan Digital Sky Survey (SDSS-IV and SDSS-V), and the Galaxy and Mass Assembly (GAMA), which offers a unique opportunity of finding these sources. In this work, we first summarize the S21 identification and classification method of 346 clusters of galaxies in disguise and further characterize the sample with SDSS follow-up observations in Section~\ref{sec:sample}. The X-ray and radio properties and assessment of the selection and completeness are provided in Section~\ref{sec:prop}. A cross-match exercise between \rosi\ and other SZ surveys is presented in Section~\ref{sec:crossmatch}. In Section~\ref{sec:agnclusters} we present galaxy clusters and groups that host bright AGN in their BCGs, which are ideal sources for further follow-up studies. Throughout this paper, we assume a flat $\Lambda$CDM cosmology with $\Omega_\textrm{m}=0.3$ and $H_0=70~\rm km~s^{-1}~Mpc^{-1}$ unless stated otherwise.

\section{Sample, source classification, and data analysis}
\label{sec:sample}

The eFEDS field was observed before the start of the \rosi\ all-sky survey. Its location is optimized to take advantage of coverage by the other multiwavelength surveys, through LOFAR in the radio, the HSC Wide area Survey \citep{Aihara2018}, the KIDS-VIKING \citep{Kuijken2019}, the Dark Energy Spectroscopic Instrument (DESI) Legacy Imaging Survey \citep{Dey2019}, GAMA \citep{Driver09}, WIGGLEz \citep{Drinkwater2018}, The Large Sky Area Multi-Object Fibre Spectroscopic Telescope (LAMOST) \citep{Zhong2020}, and SDSS \citep{Gunn1998, Blanton2017, Smee2013} in the optical wavelengths. The depth of $\sim$2.2~ks ($\sim$1.2~ks after taking the vignetting into account) of the survey slightly exceeds the depth that will be reached at the final \rosi\ all-sky survey at the equatorial region.

The X-ray data processing for the eFEDS field was extensively described in \citet{Brunner2021}. We here only provide the main steps. The analysis is performed with the latest \rosi\ Standard Analysis Software System ({\tt eSASS}, version {\tt eSSASusers$\_$201009}). The calibration, astrometric calibration, GTI, flare, and bad-pixel filtering are applied using the {\tt eSASS} algorithms to obtain clean event files for each telescope module (TM).
The main eFEDS catalog is constructed by running a sliding-cell detection algorithm using all telescope modules in the 0.2--2.3~keV band, where the \rosi\ effective area is maximum \citep[see][for details]{Liut2021}. The main source catalog, which is produced by applying a cut over the detection likelihood of 6 or higher, consists of 27910 sources. A subsample of 542 extended sources is compiled by selecting sources with a source extent larger than zero, an extent likelihood higher than 6, and a detection likelihood higher than 5 \citep{Liu2021, Klein2021}. We only use this sample for comparisons in this work, while the detailed analysis of these clusters is provided in \citet{Liu2021}, \citet{Ghirardini2021b}, \citet{bahar2021}, and \citet{Pasini2021a}. 

The point source catalog, on the other hand, is constructed from the main eFEDS catalog by applying a selection based on a detection likelihood higher than 6 and an extent likelihood of 0. This selection results in a total of 27369 point-like sources in the eFEDS field. Some of the extended sources can be mixed in the point source catalog for various reasons, including a bright central AGN, bright cool-cores, and/or faint and compact extended emission. Our first goal is to understand this selection process using the \rosi\ X-ray observations of these sources. To this end, we use the clusters identified in the point source catalog presented in S21, where the details of the optical identification and classification process are described extensively. The authors ran the red-sequence identifier code (MCMF) to find the concentrations of red galaxies around the detected point sources to locate the clusters of galaxies in the point source catalog \citep[see][for further information about MCMF]{Klein2019, Klein2021}. In this section, we summarize their point source identification and classification strategy, as well as the X-ray analysis of the detected extended sources.

\begin{figure}
\begin{center}
\includegraphics[width=0.44\textwidth]{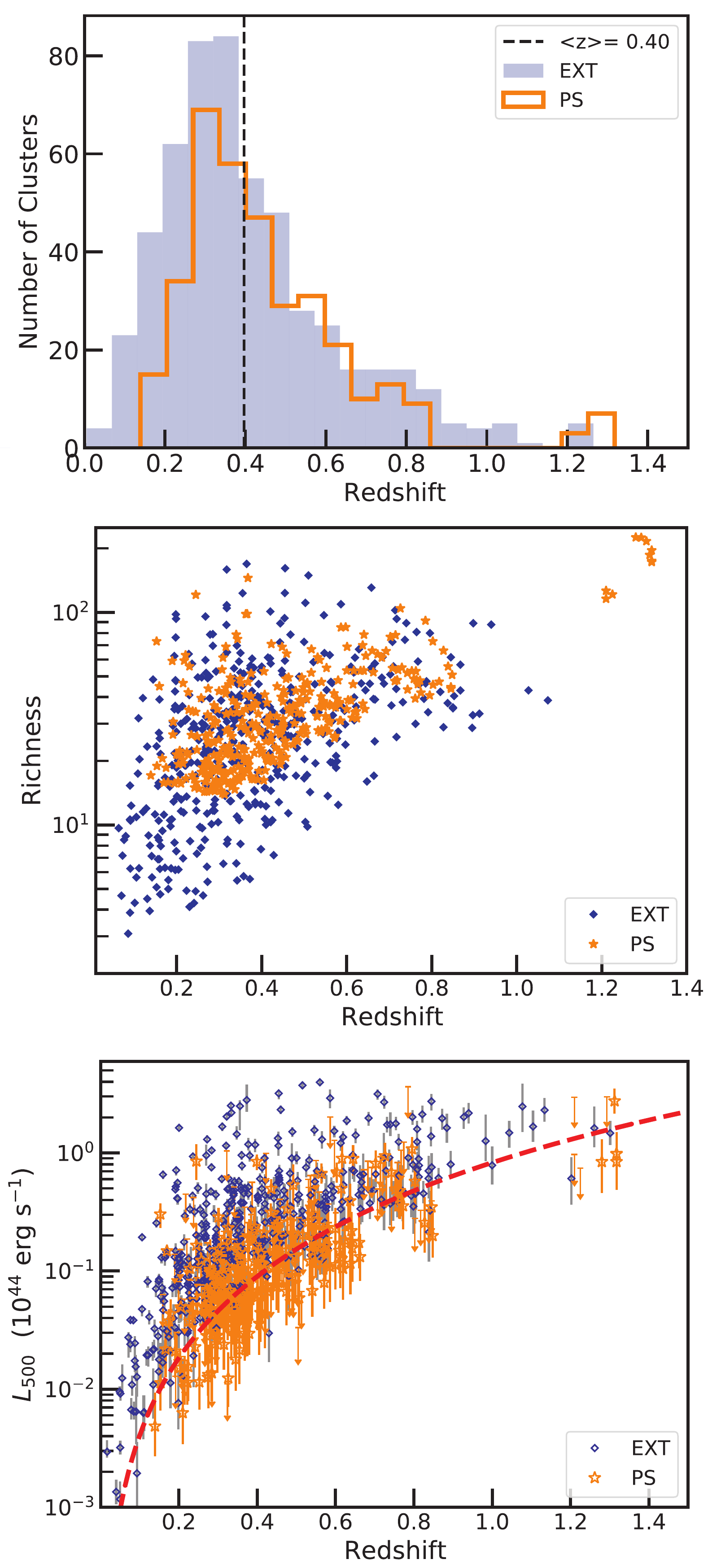}
\vspace{-1mm}\caption{Global properties of the 346 clusters and groups in the eFEDS point source catalog provided by MCMF. {\bf Top panel:} Redshift distribution of the 346 clusters and groups in the eFEDS point source catalog provided by MCMF (blue). The median redshift of this sample is 0.40, i.e., slightly higher than the redshift distribution of the extent-selected clusters in \citet{Liu2021} ($z_{med}^{ext}=0.35$; shown as gray bars). The redshift distribution of the extent-selected sample and clusters in the point source catalog is skewed toward the low redshifts, as expected.  {\bf Middle panel:}  The richness distribution of the clusters in the extent-selected sample (blue) with the clusters found in the eFEDS point source catalog (orange). Optical cleaning filtering of $f_{cont} < 0.2$, recommended by \citet{Klein2021}, was applied to both samples. {\bf Bottom panel:} The distribution of the soft-band ($0.5-2$~keV) luminosity as a function of redshift for the point source sample in orange, compared with the extent-selected sample in blue. The detections with $<2\sigma$ are shown as upper limits. The dashed red line marks the flux limit  $1.5\times10^{-14}$ ergs s$^{-1}$ cm$^{-1}$ of the extended source sample reported in \citet{Liu2021}.}
\label{fig:zhist}
\end{center}
\end{figure}
\subsection{Counterparts to the point source catalog}
The method adopted to identify multiwavelength counterparts to the point sources catalog was presented in detail in S21. In short, two independent methods ({\sc NWAY} and {\sc astromatch}) were applied and the results were compared. Both methods (the first based on Bayesian statistics and the second based on maximum likelihood) used key features typical of X-ray emitters (regardless of their Galactic or extragalactic nature). While the key features (a mixture of fluxes, magnitudes, colors, and other properties) adopted by {\sc NWAY} and {\sc astromatch} are different and identified by different algorithms, they were extracted from the multiwavelength properties of the {\it \textup{same}} training sample of about 23,000 {\it XMM} detected sources spanning the same X-ray flux of eFEDS and with reliable counterparts in the Legacy imaging survey supporting the Dark Energy Spectroscopic Instrument \citep[DESI;][]{Dey2019}. Based on these features, both methods assign the probability of being the correct counterpart to each optical source within $30\arcsec$ from the X-ray position. Then this probability is combined with the probability of a correct association based on the spatial distribution (separation, positional errors, number density). Both methods applied the procedure not only to the eFEDS sources, but also to a common validation sample of 3500 {\it Chandra} sources with secure counterparts that were made eROSITA-like in terms of positional accuracy.

Survey catalogs compiled by an automated source detection procedure can be characterized in terms of completeness and contamination because the catalogs may show spurious contaminating or missed sources. We determined the completeness level of the survey by a fraction of detected true sources in the simulations above a given flux threshold. Contamination is defined as the fraction of spurious sources, and purity is defined as the fraction of true sources in the catalog. Based on the results of the validation sample, we demonstrated that both methods perform very well. The purity and completeness exceed 95\%. {\sc NWAY} performs slightly better because it has a small number of sources with multiple possible associations. The counterparts assigned by the two methods are the same for 24193 of the 27369 sources in eFEDS (88.4\%). The number of disagreements increases with decreasing X-ray detection threshold, which suggests that some of these sources might be spurious detections. Based on the comparison of the proposed counterparts and the probability value from the two methods, we graded the reliability of counterparts (\texttt{CTP\_QUALITY}). Further details are given in S21, and we provide the \texttt{CTP\_QUALITY} of our sources in the next section.

\subsection{MCMF identification of clusters}

The multicomponent matched filter \citep[MCMF;][]{Klein2018,Klein2019} cluster confirmation tool is designed to identify ICM-selected clusters out of large samples candidates with a high degree of contamination. Its initial application on the ROSAT-based 2RXS catalog resulted in one of the largest X-ray-selected cluster catalogs to date without a selection based on source extent \citep{Klein2018}. At first glance, the application of MCMF on the eFEDS point source catalog is straightforward because the main difference is the much lower flux limit of the sample.

This impression turned out to be not true and forced changes to the default way of running MCMF, as well as to the selection of cluster candidates. In addition to the difference in flux limit,  further differences complicate the correct identification of clusters in the point source sample. The first difference is related to the resolution and source measurement. In 2RXS, a fixed and large aperture of 5\arcmin\ was used to measure the source flux, which essentially includes most of the flux for clusters with z$>$0.2. In the case of eFEDS, a source that falls below the thresholds of being extended (add extent likelihood and extent thresholds) is measured with a point-like aperture (1\arcmin) that typically just contains a fraction of the total cluster flux. Furthermore, clusters showing more complex shapes can be split into multiple sources, each containing only a fraction of the cluster flux. Because of the selection criteria of the point source sample and the depth, the fraction of clusters in that sample is even smaller than in 2RXS. Finally, the low flux limit causes a very high source density of real point sources (e.g., AGN and stars), partially projected onto clusters. 

These difficulties affect the MCMF confirmation in two ways. First, the X-ray flux from the source detection is not a good measure of the cluster flux. We therefore remeasured the richness of the candidates found by the default MCMF run using optical data alone.
Second, we excluded all candidates and randoms within a radius from the extent-selected sources for the calibration of $f_\mathrm{cont}$ based on redshift and richness. This avoids biases caused by multiply counting the same extended cluster because of proximity to the cluster or overlaps with the cluster itself. The contamination fraction, $f_{\rm cont}$, determines the probability for an optical concentration of red galaxies to be a chance alignment along the line of sight to the X-ray source. The randoms here are defined as randomized positions that exclude regions around the real cluster candidates \citep[see][for further details]{Klein2018, Klein2021}. Multiple counting of the same clusters just presented in the
point-like sample remains unaccounted for in the derivation of $f_\mathrm{cont}$. As these systems can be assumed to be smaller in extent because of the X-ray selection, the main reason for multiple counting of clusters is source splitting. Based on the additional information from the dedicated search for point-like counterparts of S21, we selected cluster candidates using the combined information of MCMF and the point-like candidates. With these selection criteria, we find 346 sources with a simultaneous concentration of red-sequence galaxies \citep[see][for details]{Salvato2021}. The locations of these clusters in the eFEDS field are shown in Figure~\ref{fig:cat}. The full MCMF catalogs typically include the information about the BCG positions, optical centroids, $f_\mathrm{cont}$, an optical mass proxy, photometric redshifts, and richness \citep[see][for details]{Klein2021}. In this catalog, we only provide MCMF redshifts and richness because the remaining sample properties are provided in \citet{Salvato2021}.

In Figure~\ref{fig:zhist} we provide the redshift histogram of the full sample in the point source catalog. Compared to the extent-selected sample (z~$_{\rm med}=0.35$), the median redshift of the sample is slightly higher (z~$_{\rm med}=0.40$). This is expected because the apparent size of the cluster should be smaller than the sizes in the extent-selected sample for them to be missed by our selection criteria, which are based on the extent of the source. We identified an additional 18 clusters at high redshifts (z~$>0.8$), 10 of which lie at redshift z~$>1$. The middle and bottom panels of the same figure compares the richness and luminosity measurements of the clusters in the point source sample (in orange). The richness and luminosity of this sample is similar to the extent-selected sample, which is marked in blue. 

 The clusters and groups in the point source catalog are classified in S21 based on their counterparts in the optical data in the following scheme (in the order of cluster likelihood):

{\bf Class 5} marks the red-sequence galaxy concentration detected significantly with $f_{cont}<0.2$ by MCMF in proximity of the X-ray position for which there is no secure point source counterpart in S21 (CTP$\_$QUALITY$\le$1). The probability of the source being in a cluster is $f_{cont}<0.2$. These clusters could be missed by the extent of the likelihood cut. This category includes 120 clusters or groups in the catalog. 

{\bf Class 4} sources are the clusters that have secure optical counterparts from S21 with CTP$\_$QUALITY$\ge$2 that the optical colors of the counterpart are consistent with a red galaxy member of the cluster identified by MCMF with a high  $f_{cont}<0.2$ probability at the same redshifts. The sources also lie in the elliptical locus in the z-W1 vs g-r plane; see figure 21 in S21. The BCGs identified by both MCMF and NWAY are the same passive red galaxy. We find 63 clusters that fall in this category. 

{\bf Class 3} consists of clusters that correspond to a secure counterpart listed in S21 with CTP$\_$QUALITY$\ge$2 and MCMF with $f_{\rm cont}<0.2$. The redshift from LePhare of the identified point source counterparts and the redshift of the cluster found by MCMF agree. Unlike Class~4, the colors are more typical of AGNs, QSOs, or a hybrid (galaxy + AGN template; see figure 21 in S21). These clusters consist of AGN(s) in their vicinity that are associated with the X-ray detection. We find 96 cases in the catalog. 

{\bf Class 2} point source candidate is identified by S21 with high probability (CTP$\_$QUALITY$\ge$2) and has a different redshift than the cluster identified by MCMF. The cluster red sequence is also detected with a very high probability of $f_{\rm cont}<0.01$. These cases can be interpreted as point sources that are in projection with clusters on the line of sight. We find 67 cases in total in this group. We note that the threshold likelihood applied to MCMF is more stringent than the classes above, and therefore there is a clear association with a cluster. For the point sources near the X-ray position, further analyses with instruments with high spatial angular resolution are required to determine the fraction of X-ray emission that is associated with the cluster or the point source counterpart.

The redshift and richness distribution of the sample shows no immediate correlation between these observables and the $Cluster\_Class$ (Class, hereafter). Because the clusters and groups in this sample have a relatively high probability ($>80$\%; $f_{\rm cont}<0.2$) of being secure extended sources, we assume that they are all genuine clusters and groups throughout the paper. 

We note that we applied a richness cut due to the optical cleaning applied ($f_{cont}<0.2$) for both extended and point source samples. The clusters and groups in this sample have a relatively high probability ($>80$\%) of being secure extended sources.

\subsection{\rosi\ X-ray data analysis}
\label{sec:xrayanalysis}
The method of X-ray analysis is very similar to the analysis of the extent-selected sample, and the details were already provided in \citet{Ghirardini2021b, Ghirardini2021a}. We therefore only summarize the main steps of the X-ray analysis. The imaging analysis was performed in the soft $0.5-2$~keV energy band using the {\tt eSASS} (version {\tt eSASSusers\_201009}) tools {\tt evtool}, and the exposure maps were created using {\tt expmap} surrounding the centroid of the point sources. The images and exposure maps with $15^{\prime}$ at a side were created for further processing.  The point sources within the field of view were either modeled or excised from images based on their flux \citep[see][for further details]{Ghirardini2021a, Ghirardini2021b}. The analysis procedure was applied to the extended source analysis, but in this sample, we did not excise the center from the analysis, neither do we provide core-excised properties in this work. We fit these images with a total image model after deprojection of the model to the image plane. The model included a cluster component based on the \citet[][V06, hereafter]{vikhlinin2006} formulation. Our background model consisted of vignetted Galactic soft thermal foreground model, an absorbed power-law model for the cosmic X-ray background component due to unresolved sources, and an unvignetted instrumental background model that was based on the best-fit model of the filter-wheel-closed data \footnote{https://erosita.mpe.mpg.de/edr/eROSITAObservations/EDRFWC}. We excluded the point sources within the fitting radius by modeling them as delta functions convolved with the point spread function (PSF) of \rosi. The model gives the best-fit parameters of the V06 density model, which can be converted into a gas density by $\rho_{g} = \mu_{e}n_{e} m_{p}$, where $\mu_{e}$ is the mean molecular weight, $n_{e}$ is the number density of electrons, and  $m_{p}$ is the proton mass \citep{bulbul2010}.
\begin{figure*}[ht!]
\begin{center}
\includegraphics[width=0.49\textwidth]{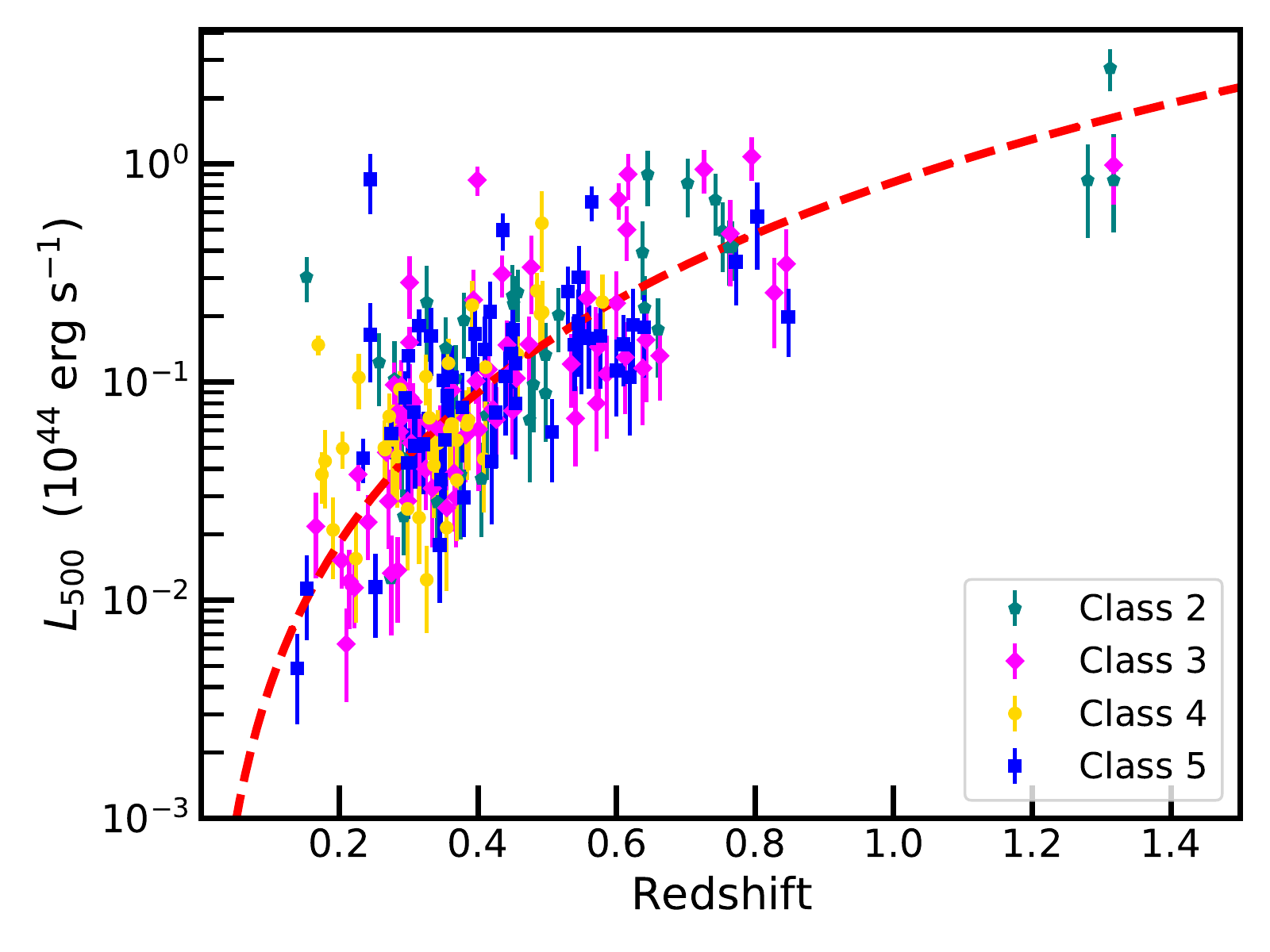}
\includegraphics[width=0.49\textwidth]{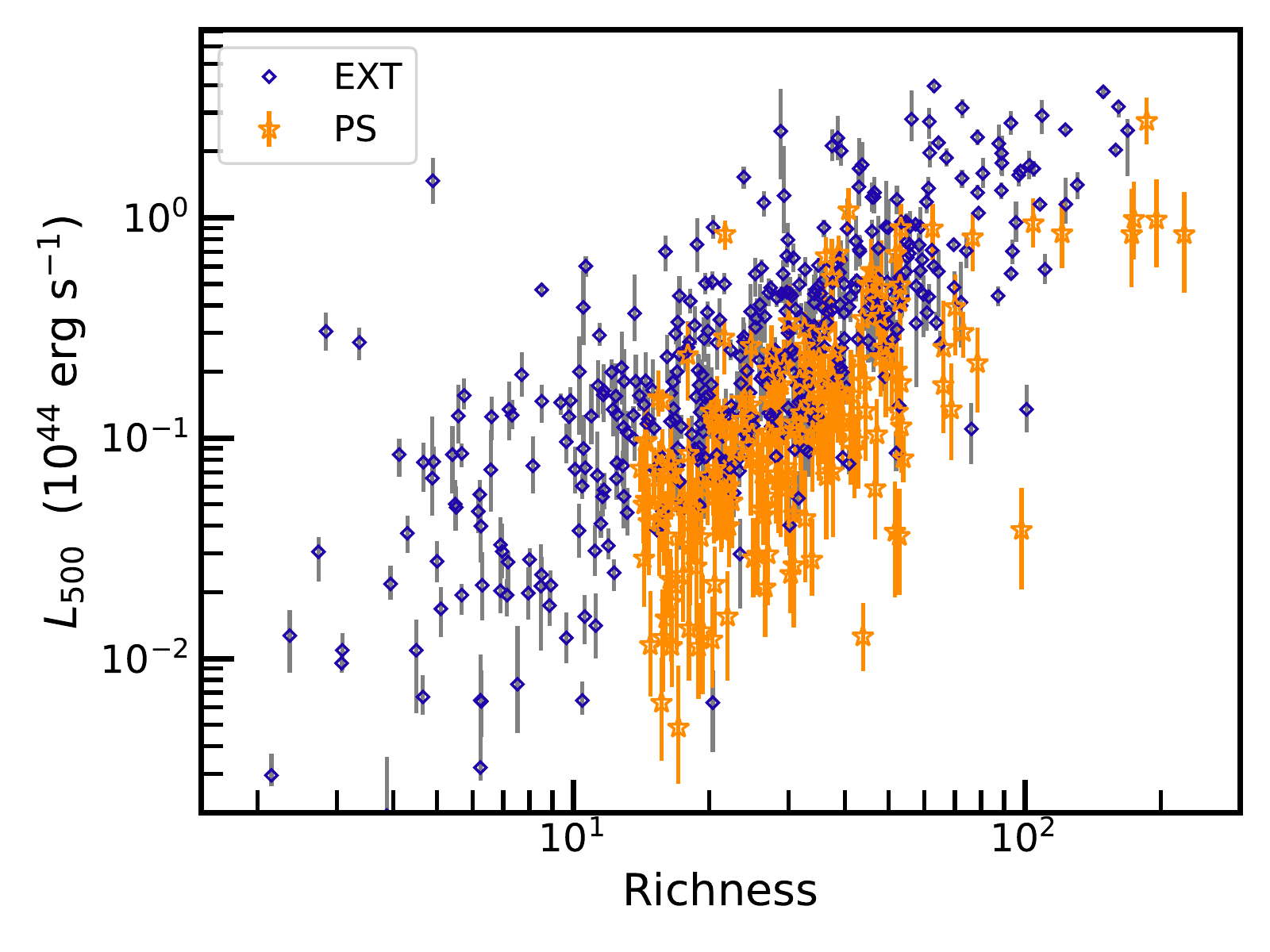}
\caption{Luminosity distribution of clusters with respect to redshift, class, and richness. {\bf Left panel:} Luminosity-redshift distribution of the clusters and groups identified in the point source sample based on their class. A 2$\sigma$ detection limit is applied on the luminosities. The flux limit for the extent-selected sample of 1.5$\times10^{-14}$ ergs s$^{-1}$ cm$^{-2}$ is displayed as the dashed red line \citep{Liu2021}. The luminosity of most clusters in the point source catalog is lower than that of the extent-selected sample, indicating that this sample is composed of clusters and  groups that are smaller in extent and lower in mass, whose emission might be dominated by the central AGN.  {\bf Right panel:} X-ray luminosity as a function of the red-sequence richness. The blue data points show the extent-selected sample \citep{Liu2021, Klein2021}. The orange points mark the clusters in the point source sample. }
\label{fig:lumin}
\end{center}
\end{figure*}

The spectra and corresponding response files were extracted using the {\tt eSASS} tool {\tt srctool} within the overdensity radius $R_{500}$. The overdensity radius $R_{500}$ is the radius within which the local density of the cluster is 500 times the critical density of the Universe at the cluster redshift. The overdensity radii $R_{500}$ of these clusters were determined using the best-fit luminosity - mass scaling relations of the extent-selected sample based on the HSC weak-lensing observations \citep{chiu2021, bahar2021}. The background spectra were extracted between 4--6~$R_{500}$ surrounding the cluster centroid. We fit the spectra using the software package {\it XSPEC} version 12.11.1 \citep{arnaud1996}. The ICM emission was modeled with the soft thermal absorbed {\it apec} component \citep{smith2001,foster2012}, and the local background was modeled with two components for the emission from the Local Hot Bubble (or heliosphere); a cool unabsorbed single-temperature thermal component (E$\sim$0.1~keV), an absorbed thermal component (E$\sim$0.2~keV) for the Galactic hotter halo and intergalactic medium, and a nonthermal component for the unresolved cosmic sources as described in \citet{bulbul2012, Bulbul2019}. The particle background is also added to the total background model. In this case, the model parameters were determined from the best-fit values of the shape parameters of the filter wheel closed data and are kept frozen \citep{freyberg2020}, while the normalizations were left free in the fit. We adopted the \citet{asplund2009} table for the abundances and the {\tt tbabs} \citep{wilms2000} model with \citet{willingale2013} for the column density $n_{\rm H}$ to account for the local absorption. Because most of the detected clusters lie are in the low-count regime, we strictly used C-statistics in our fits \citep{cash1979, kaastra2017}. In the fits, the temperature and normalization of the cluster model were left free, while the metallicity was fixed. Considering their smaller extent and low flux, it is not surprising that we cannot reliably measure the ICM temperature or abundance for most of these clusters. Additionally, because the PSF size of 26$^{\prime\prime}$, it is challenging to excise the central AGN when the spectra are extracted. When the central core is excised, we fail to measure the ICM temperature of the majority of these clusters because the count statistics are too low. We therefore adopted a uniform approach and fixed the abundance value to 0.3 solar abundance, and for the temperature, we assumed a probability distribution identical to the one observed in the eFEDS cluster catalog  \citep{Liu2021}. The temperature of a cluster was randomly drawn and allowed to vary smoothly around $2.4_{-0.9}^{+1.4}$~keV, where 2.4~keV is the median temperature, and 0.9~keV and 1.4~keV are the differences between the median and the 16$^{\rm th}$ and 84$^{\rm th}$ percentiles of the eFEDS cluster temperatures, respectively.
These values of temperature and abundance were used 
in the conversion of the count-rate and surface brightness into the soft band $0.5-2$~keV luminosity. To further calculate the masses of these clusters, we calculated the total masses based on the scaling relation of the luminosity to weak-lensing mass of the extent-selected sample presented in \citet{chiu2021}. The X-ray properties of the clusters and groups in this sample are provided in Table~\ref{tab:main}.

\section{Characterization of the sample and classification with multiwavelength observations}
\label{sec:prop} 

To be able to characterize the properties of the sample and understand the selection, completeness, and contamination of the extended sources in the point source sample, we furthermore used the \rosi\ X-ray data, SDSS optical spectroscopy, and LOFAR radio observations of the eFEDS field. Our goal is to understand the reasons for the mischaracterization of the extended sources by the source detection algorithm and contamination in the sample, and to test the classification by comparison with the extended source catalog and through available multiwavelength observations.

\begin{figure}[ht!]
\begin{center}
\includegraphics[width=0.49\textwidth]{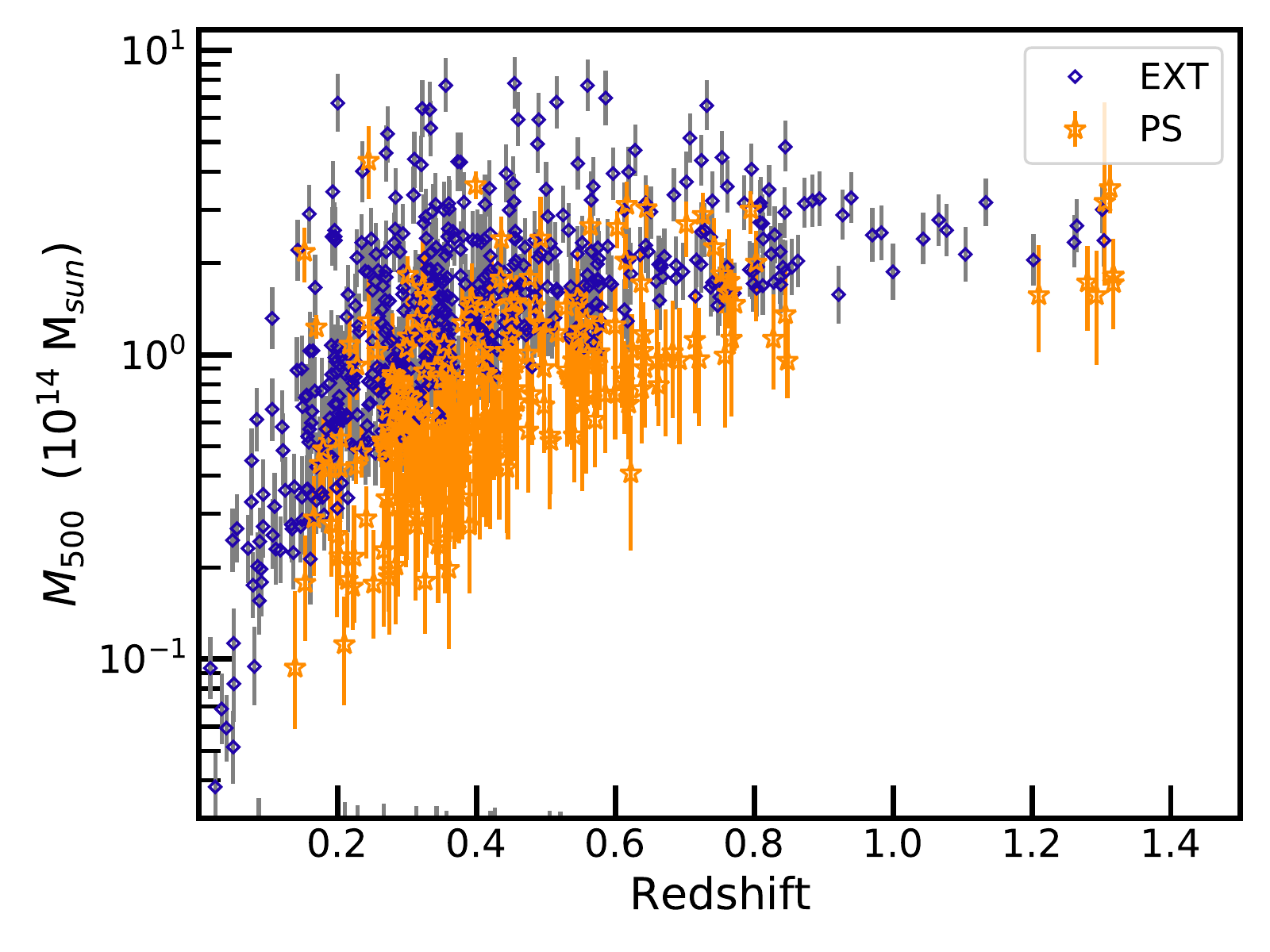}
\caption{
Mass-redshift distribution of the clusters in the extent-selected sample (blue) and the clusters identified in the point source sample (orange). The clusters found in the point source sample are fainter than the extent-selected companions in the eFEDS field, suggesting that they are mostly missed due to our extent or detection likelihood selections. 
}
\label{fig:mass}
\end{center}
\end{figure}

\subsection{Characterization of the X-ray selection through \rosi\ observations}
\label{sec:xrayprop}

The luminosity measurements within $R_{500}$ of the full sample are shown in the left panel of Figure~\ref{fig:lumin}. Because of the sizable PSF of \rosi\ \citep[$\approx 26^{\prime\prime}$ HEW FOV average;][]{predehl2021} and the small angular scales of the clusters in this sample, it is very challenging to excise the cluster core, in which the core emission dominates the overall emission in determining the core-excised luminosities. Therefore, when an AGN is present in the core or in projection (e.g., in Class~2 and 3 clusters), the flux and luminosity measurements reported in this work include the no-thermal X-ray emission from the central region and can be biased high. The measurements presented here should be taken as approximate upper limits of the luminosity. A follow-up study with deeper and higher spatial resolution X-ray observations with \xmm\ and \chandra\ will help us to recover the unbiased ICM luminosity and temperatures of these clusters and groups. In Figure~\ref{fig:lumin} we present a comparison of luminosity measurements with a significance of $>2\sigma$. Figure~\ref{fig:lumin} shows that a large fraction of the clusters (113, to be precise, corresponding to 33\% of the sample) lie below the flux limit of 1.5$\times10^{-14}$ ergs~s$^{-1}$~cm$^{-2}$ of the extent-selected sample \citep{Liu2021}. X-ray emission from a subset of these clusters is also likely to be boosted by the AGN emission, where it is present. This boosting effect would push the luminosities above the detection limit of the eFEDS point source catalog. In the absence of the AGN (at the center or in projection), these sources would lie beyond the eFEDS survey detection limit and would only be detected with deeper surveys. The class distribution of the eFEDS point-like clusters does not show a clear trend with luminosity. Hence, we cannot make a firm conclusion about the nature of the sources and why the clusters in different classes are detected in the point source sample.

We also examined the richness and X-ray luminosity distribution of the clusters of the sample, shown in the right panel of Figure~\ref{fig:lumin}. The clusters that have high X-ray luminosities for their richness may be those that host a bright quasar in their BCG. However, we note that optical richness alone is not a reliable proxy for the cluster luminosity because of the large scatter due to projection effects \citep{Ge2019}. 

A consistent finding is also reflected in the total mass distribution of the sample, where the majority of these sources are in fact low-mass clusters or groups (see Figure~\ref{fig:mass}). Because the extent-selected clusters are a representative sample of clusters above the given flux limit, we used the \citet{chiu2021} mass scaling relations, constructed from the soft-band X-ray luminosity and HSC weak-lensing mass measurements of the extent-selected clusters, to calculate the masses of the clusters of galaxies in the point source sample. The masses of most clusters lie between $1\times10^{13}$~$M_{sun}$ and $3\times10^{14}$~$M_{sun}$, which is lower than the extent-selected sample. We note that the masses and luminosities reported here for Class~2 and 3 objects should be taken as upper limits as these clusters either have an AGN in projection or an AGN in their BCGs.
\begin{figure*}[ht!]
\begin{center}
\includegraphics[width=0.49\textwidth]{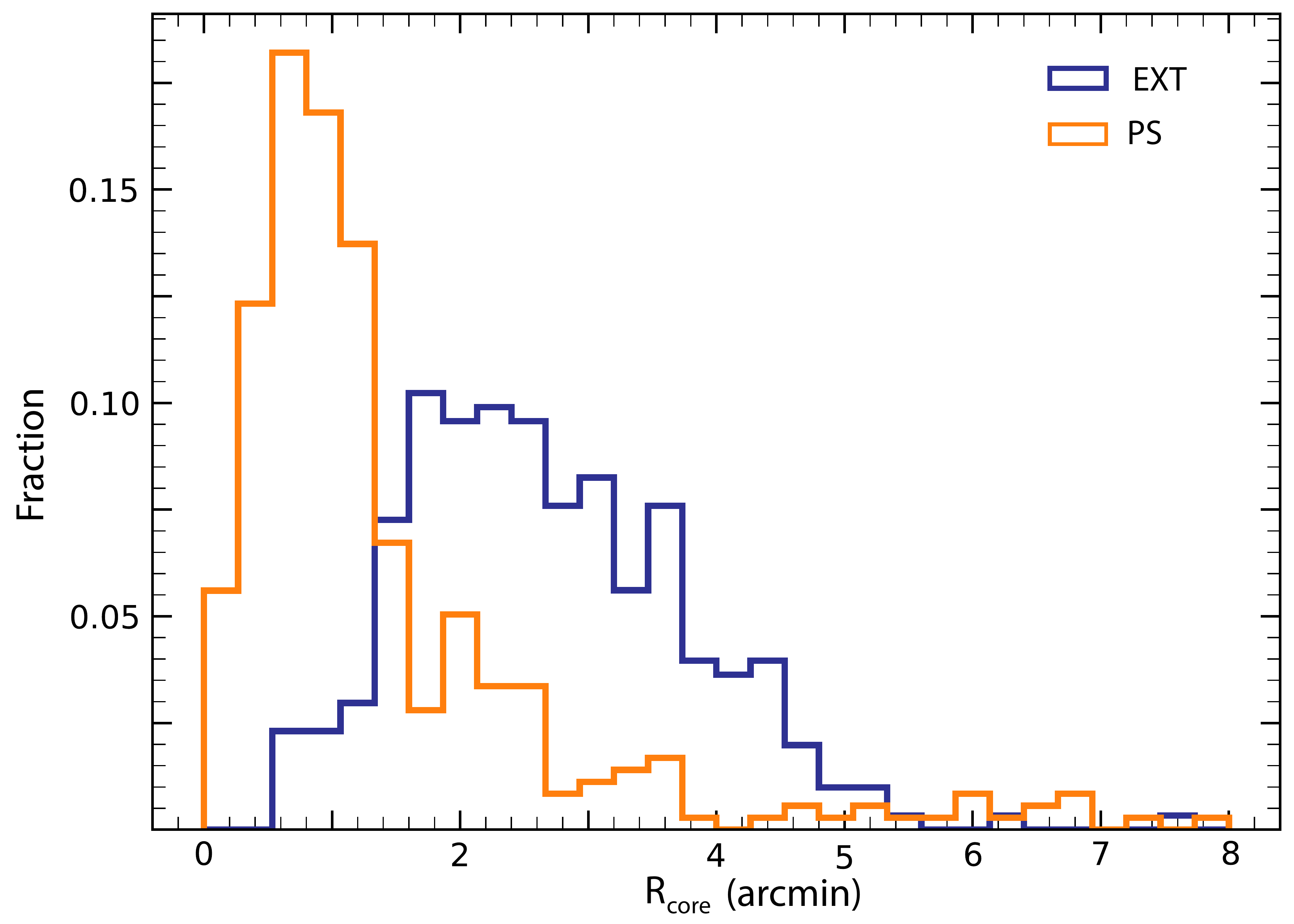}
\includegraphics[width=0.49\textwidth]{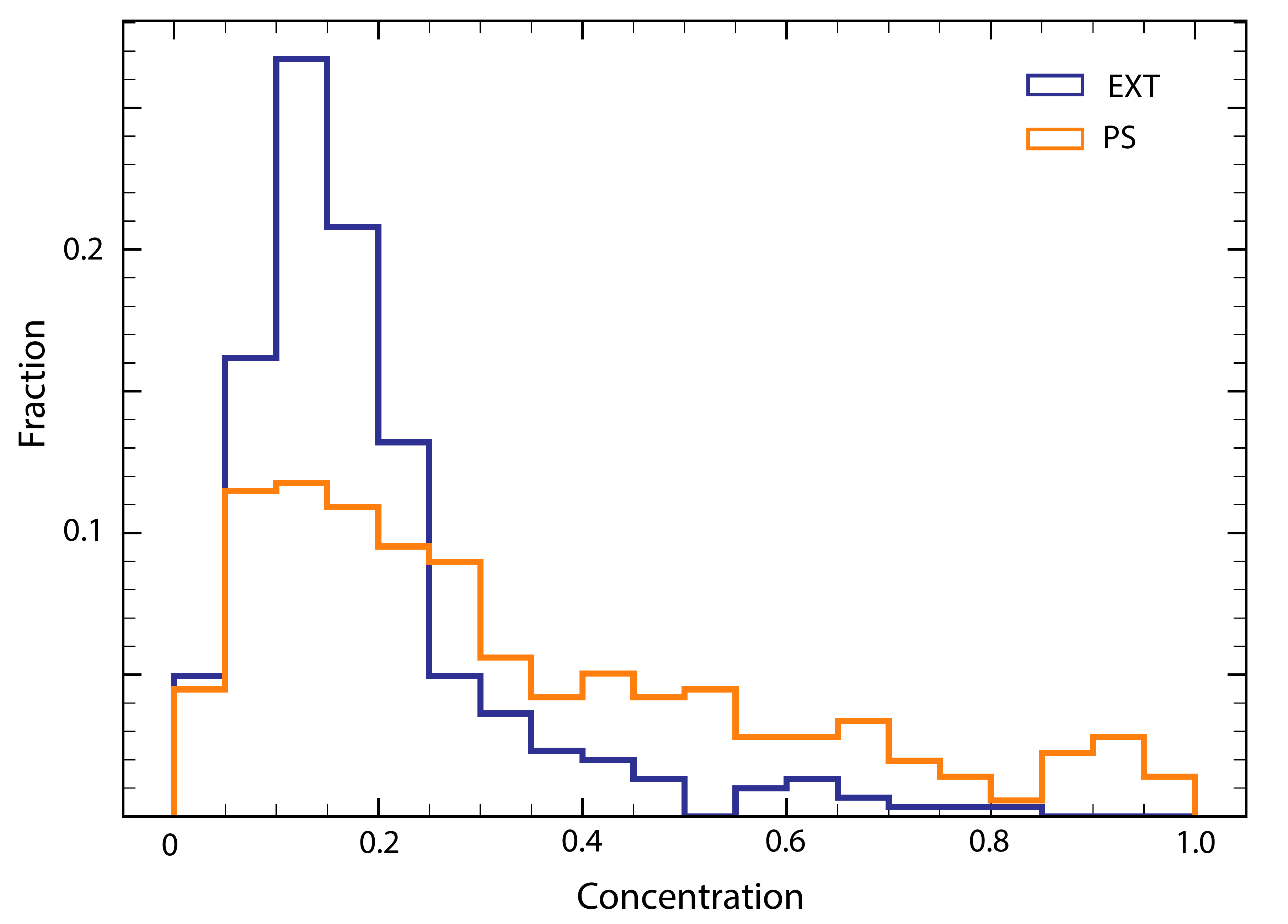}
\caption{Distributions of the $R_{core}$  and $c_{\rm SB}$ as a measure of compactness of the clusters in the point source catalog and the extent selected catalogs. The left panel shows the distribution of $R_{core}$ of the cluster sample in the point source catalog (orange) and the extent-selected ($L_{\rm ext}>12$) cluster sample (blue). The right panel shows that the distribution of the $c_{\rm SB}$ of the clusters in the point source catalog is smaller than in the extent-selected sample, indicating that the clusters in the point source catalog are more compact.   }
\label{fig:rccsb}
\end{center}
\end{figure*}

It is also possible that these clusters have a smaller extent and can just be missed by our extent selection as our detection algorithm sets the extent to zero if it is smaller than 6 \citep{Brunner2021}. Following the method presented in \citet{Ghirardini2021b}, we estimated several dynamical properties of the clusters in the point source sample and compared them with the extent-selected sample presented in \citet{Ghirardini2021b}. In Figure~\ref{fig:rccsb} we compare the distributions of the core radii ($R_{core}$) constrained by the V06 model and the concentration parameter ($c_{\rm SB}$)  between these two samples. The concentration parameter is defined as the ratio of the surface brightness within 0.1R$_{500}$ to the surface brightness within R$_{500}$ \citep{Ghirardini2021b, Santos2008, Maughan2012}. Intuitively, the expectation is that the smaller the core radius, the more compact the cluster. The left panel of Figure~\ref{fig:rccsb} clearly shows that the clusters in the point source sample have relatively smaller core radii, hence the emission is more concentrated in a smaller area. Consistently, the concentration of the point source sample shows a clear excess in higher values than the extent-selected sample, indicating that a significantly larger fraction of cool-core clusters and clusters host a central AGN. We performed the same experiment by applying cuts in flux $1.5\times10^{-14}$~ergs~s$^{-1}$~cm$^{-2}$ and in detection likelihood to test whether the clusters are missed by the extent selection because they are fainter and/or more compact than the extent-selected clusters. The distribution of number density, core radius, and concentration parameter remains the same, indicating that the population of clusters in the point source catalog is more compact than the extent-selected sample. The extent-selected sample does not show a clear bias toward cool-core clusters or clusters with a central AGN, but contains the fraction of cool-cores is similar to that of SZ surveys \citep{Ghirardini2021b}. In this sample, we observe the opposite trend.

\begin{figure}
\begin{center}
\includegraphics[width=0.49\textwidth]{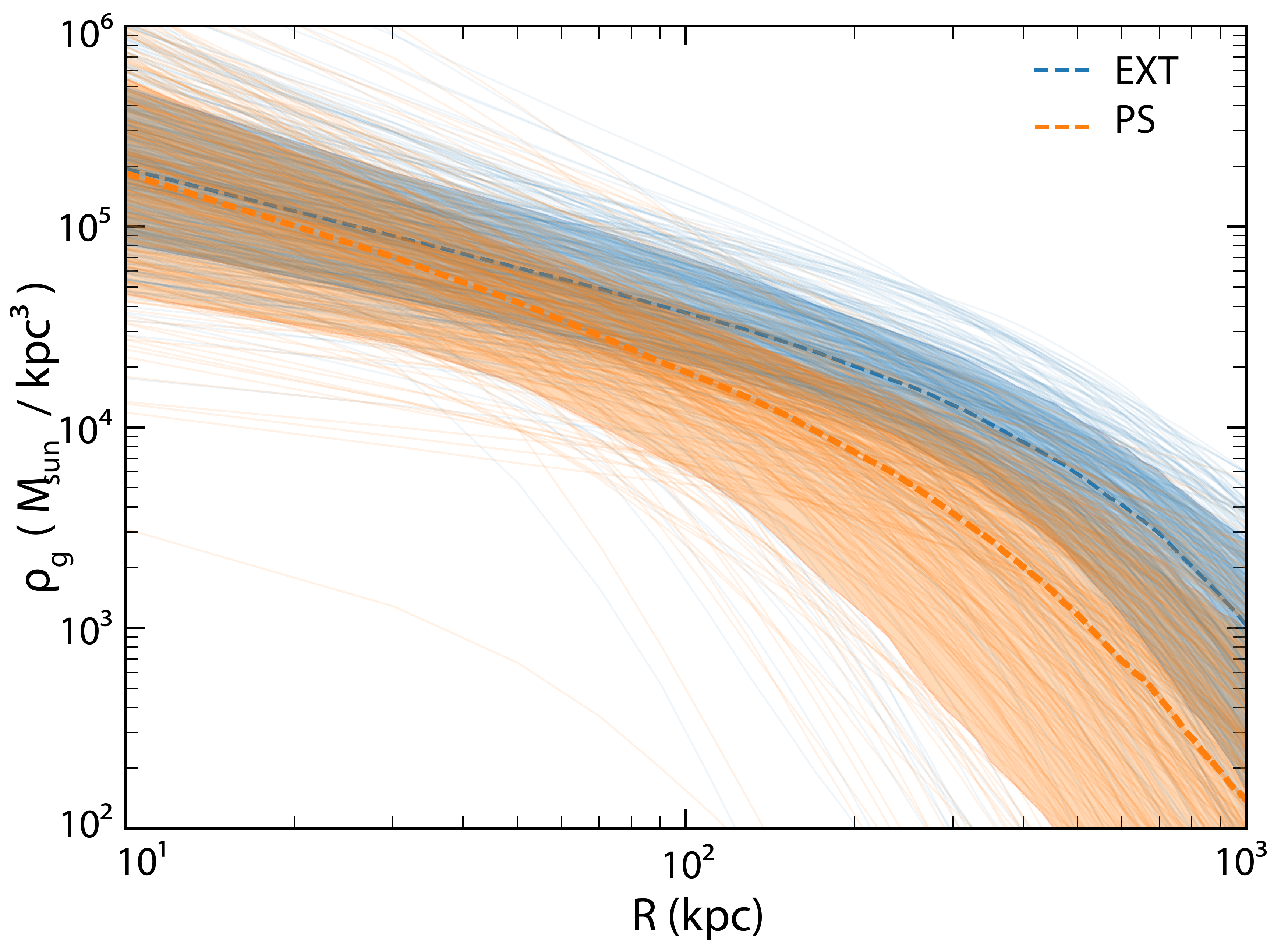}
\caption{Gas density of the clusters in the point source catalog (orange) compared with those in the extent-selected clusters (blue). The dashed curves display the median profiles of both samples. The shaded regions show the 68\% confidence intervals around the median values of the radial profiles. The selection of an extent likelihood higher than 12 is applied to the extent-selected sample to reduce the AGN contamination in the sample \citep[see][]{Liu2021, Ghirardini2021b}. The median gas density of the clusters in the point source catalog is more compact than that in the extent-selected sample. }
\label{fig:density}
\end{center}
\end{figure}
 A comparison of the gas density of the clusters in the point source catalog and extent-selected clusters is shown in Figure~\ref{fig:density}. The gas densities of the clusters in the extended source catalog in blue show an excess of X-ray emission at large radii $R>100$~kpc compared to the clusters in the point source catalog. This result confirms our earlier finding that the clusters in the point source catalog are more compact than the extent-selected clusters. This is expected because the some clusters in the point source sample are expected to be dominated by the sources with a bright AGN or a cool-core at their centers. Regardless, it is important to note that we still detect the extended emission in the clusters and groups in the point source catalog beyond $R>100$~kpc.

We also note that because AGNs are variable sources, the variability in AGN flux would also affect the detection and classification of the cluster. For instance, the decrease in the flux of the AGN in an AGN-dominated cluster would lead to a higher extent likelihood due to the relatively brighter ICM emission. The cluster is then likely to be classified as an extended source and would not be included in this sample. Moreover, the surface brightness concentration in this case decreases and becomes closer to the extent-selected sample shown in Figure~\ref{fig:rccsb}.

\subsection{Characterization of the source classification and contamination through SDSS and GAMA observations}
\label{sec:sdss}
We visually inspected the optical spectra of the counterparts, where available, for all classes of clusters to gain insights into the defined cluster classes (classes 2, 3, 4, and 5). The spectroscopic data from the GAMA DR3 or SDSS DR16 surveys \citep{Baldry2018MNRAS.474.3875B,Ahumada_SDSSDR16_2020ApJS..249....3A} were used. 
We additionally used spectra from special eFEDS plates acquired at the end of the SDSS-IV survey (\citealt{Blanton2017}, Merloni et al. in prep.). In this section, we describe the inspection of the optical spectra of these candidate clusters and groups and test our contamination level and classification. 

All BCGs identified by MCMF for which a spectrum is available look like passive galaxies. The spectrum of one BCG (eFEDS J092121.7+033050,  SRC\_ID=15322) shows signs of recent star formation, that is, (weak) narrow emission lines possibly coming from HII regions. 
For a fraction of the BCGs identified by MCMF, stellar mass estimates are available \citep{Taylor2011MNRAS.418.1587T,Comparat2017arXiv171106575C}, and all are above $10^{11}M_\odot$ , which is consistent with massive galaxies. 
Overall, the galaxy population indicated by MCMF follows expectations \citep[e.g., discussion by ][]{Clerc2020}. 

The counterparts identified by NWAY are more varied. 
Table~\ref{tab:VI:NWAY:spectra} lists the counterparts for each class split into categories based on the optical spectra: red-sequence galaxy, galaxies with emission lines, type 1 and type 2 AGN, stars, or other (the signal-to-noise ratio of the spectra is too low to decide about a classification). 
After visual inspection of the images (in addition to the spectra), we found that in some cases, the sources targeted for spectroscopy are a blend of a red compact object and an elliptical galaxy. These need more investigation to be classified in a clear  way. We leave a detailed description of these cases for future studies. They are reported in the table below.
We have optical spectra for a fraction of 13.4\%\ (47.9\%, 65.1\%, and 7.5\%) of sources from class 2 (classes 3, 4, and 5). Because of the optical selection function, this sample of spectra is not a fair sample in that it is biased toward the brighter objects. The completeness of the targeted objects is about 85\% (50\%) for an i-band magnitude of 21 (22).

For Class 4, we find that almost all spectra are passive galaxies (one of them is an AGN, one has a low signal-to-noise ratio). The misclassification rate in this class therefore appears to be negligible. For Class 5, the situation is similar. 
For Class 2 and 3, the population of counterparts found by NWAY is more varied. In Class 2 (3), 1/9=11.1\% (17/46=36.9\%) are passive galaxies, while 7/9=77.7\% (22/46=47.8\%) of the total sources have an AGN counterpart. Only 10\% of the AGN are located at the same redshift (within $\Delta z/(1+z)<$0.005) of host clusters.
These results support the classification scheme used in S21. 
We extrapolated these percentages to the complete classes to obtain an estimate of the total number of clusters in the point source catalog as $120 + 63\times0.95 + 96\times0.37 + 67\times0.11 = 222$. This represents $1\pm0.3\%$ of the point sources ($\sim 27,000$ point sources in total). With 2RXS/SPIDERS, \citet{Comparat2020AA636A97C} found that the cluster contamination of the point source catalog was about 3\%. 
With \rosi, this figure improved by a factor of 3. 

\begin{table*}
    \caption{Visual inspection classification of NWAY counterparts using optical spectra.}
    \label{tab:VI:NWAY:spectra}
    \centering
    \begin{tabular}{l c c c c c c c }
    \hline
\hline
Cluster Class                & 2      & 3      & 4   & 5 & Total    \\
Number of Clusters in Class  &  67 & 96 & 63 & 120 & 346 \\
NWAY CTP has Spectrum         & 9 & 46 & 41 & 9 & 105 \\
Fraction with spectrum        & 13.4 & 47.9 & 65.1 & 7.5 & 30.3 \\
\hline
 Red sequence galaxy       & 1  & 17  & 39  & 9 & 66 \\
 Galaxy with emission lines& 0  & 0   & 0   & 0 & 0 \\
 Type 1 AGN                & 6  & 16  & 1   & 0 & 23 \\
 Type 2 AGN                & 1  & 6   & 0   & 0 & 7 \\
 Star                      & 0  & 5   & 0   & 0 & 5 \\
 Other                     & 1  & 4   & 1   & 0 & 6 \\
 Any of the classes        & 9  & 46  & 41  & 9 & 105 \\
\hline 
\end{tabular}
\end{table*}
\subsection{Characterization of the source classification through LOFAR radio properties }
\label{sec:radioprop}
In this section, we summarize the LOFAR \citep{vanHaarlem2013} radio properties of the clusters and groups in the point source catalog and provide a comparison with the extent-selected sample. The details of the radio data analysis are already provided in detail    in  \citet{Pasini2021a}. We therefore omit them here and provide the results. 

\citet{Pasini2021a} (hereafter P21) performed a detailed study of the radio galaxies hosted in the BCG of the 542 eFEDS secure clusters \citep{Liu2021}.
We used the same radio source catalog, obtained from the LOFAR $120-168$~MHz observations covering the entire eFEDS field, to carry out a similar analysis for the clusters in the point source catalog. A detailed explanation of the observations and related radio catalog is presented in P21. Here, we just provide a brief description. Radio sources are detected on a $\sim$135$\,\mu\,{\rm Jy\, beam_{(8\arcsec\times9\arcsec)}^{-1}}$-noise mosaic with a detection threshold of $5\sigma$ using  \texttt{PyBDSF}\footnote{\url{https://www.astron.nl/citt/pybdsf}} (Python Blob Detector and Source Finder). Flux densities of the LOFAR-detected sources are 10\%\ higher on average than those in the TGSS-ADR1 (TIFR GMRT Sky Survey - Alternative Data Release 1) 150~MHz observations \citep{Intema2017}. The astrometric offsets in R.A. and Dec. between the LOFAR and FIRST (Faint Images of the Radio Sky at Twenty-cm; \citealt{Becker1995a}) detected sources are $0.13\arcsec$ and $0.04\arcsec$, respectively. The corresponding standard deviations are $0.70\arcsec$ and $0.82\arcsec$. 

\begin{figure*}
\begin{center}
\includegraphics[width=0.43\textwidth, trim=0 5 30 50, clip]{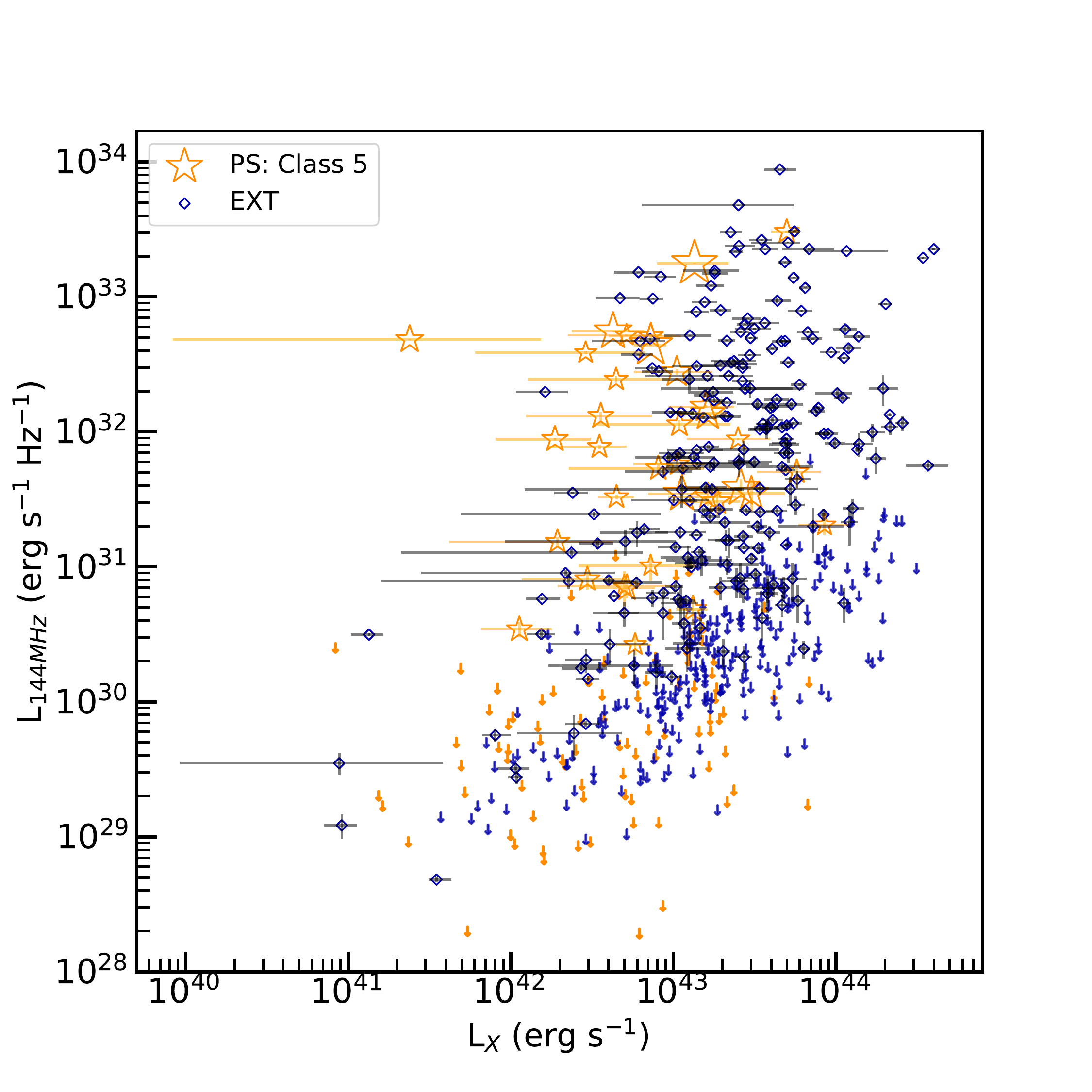}
\includegraphics[width=0.43\textwidth, trim=0 5 30 50, clip]{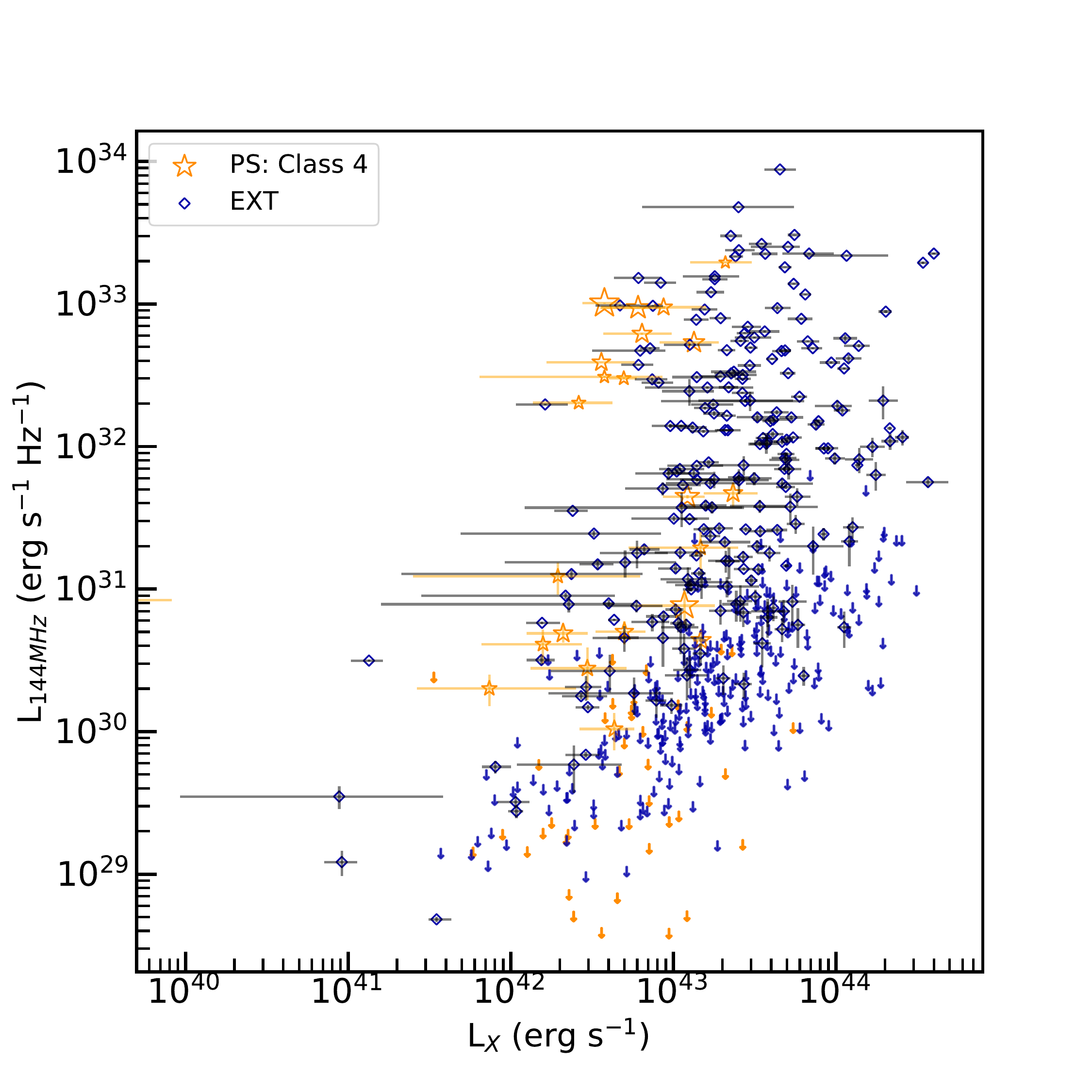}
\includegraphics[width=0.43\textwidth, trim=0 5 30 50, clip]{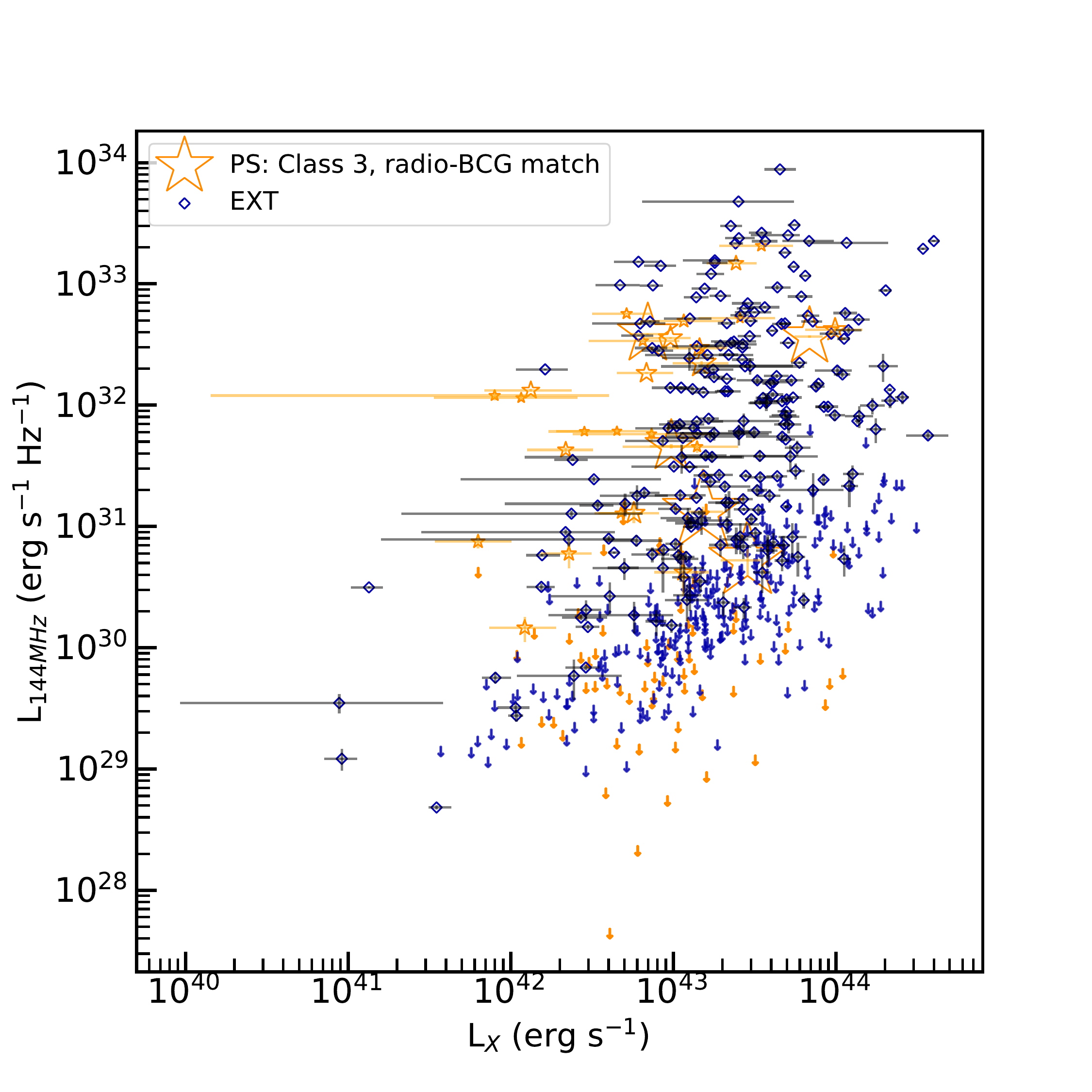}
\includegraphics[width=0.43\textwidth, trim=0 5 30 50, clip]{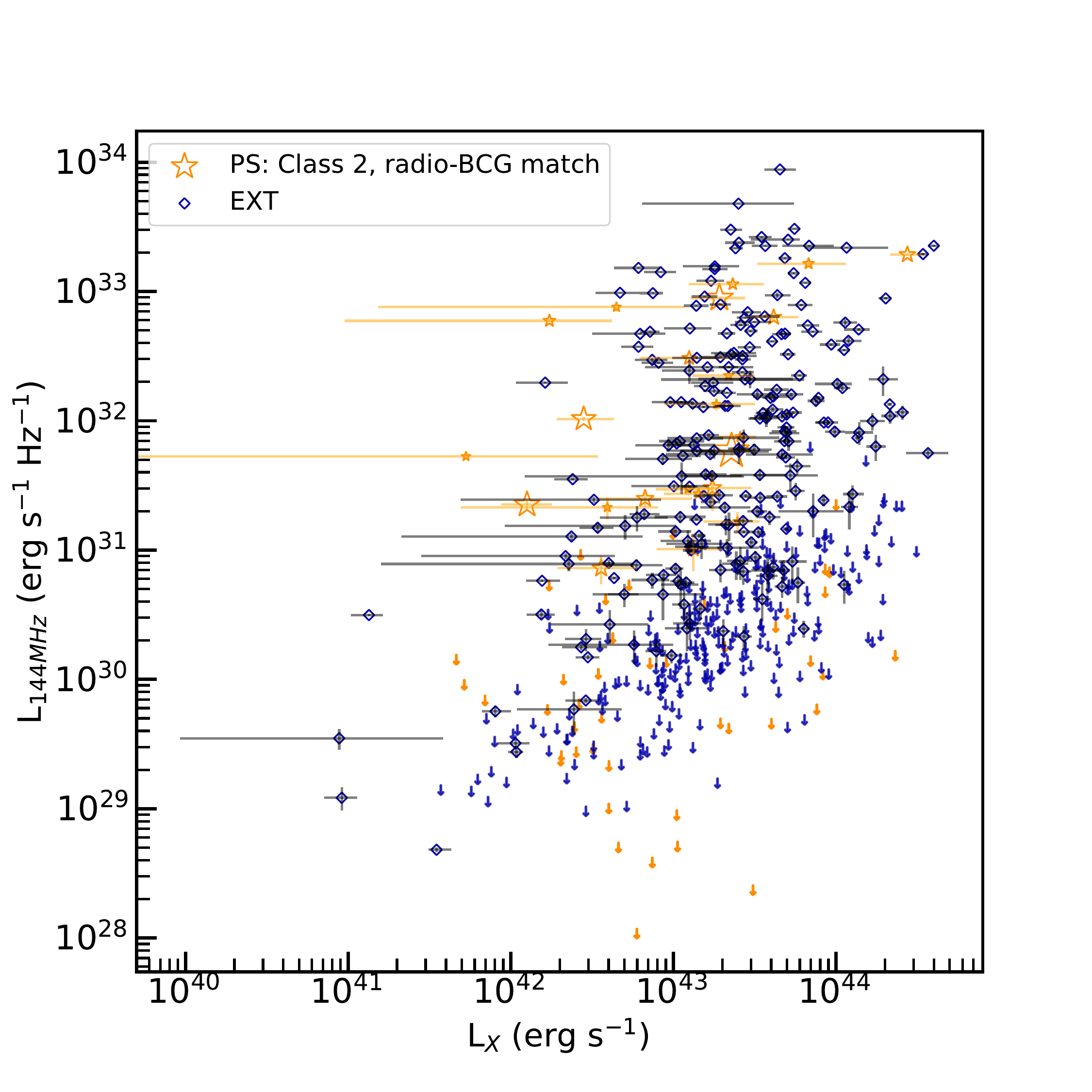}
\includegraphics[width=0.43\textwidth, trim=0 5 30 50, clip]{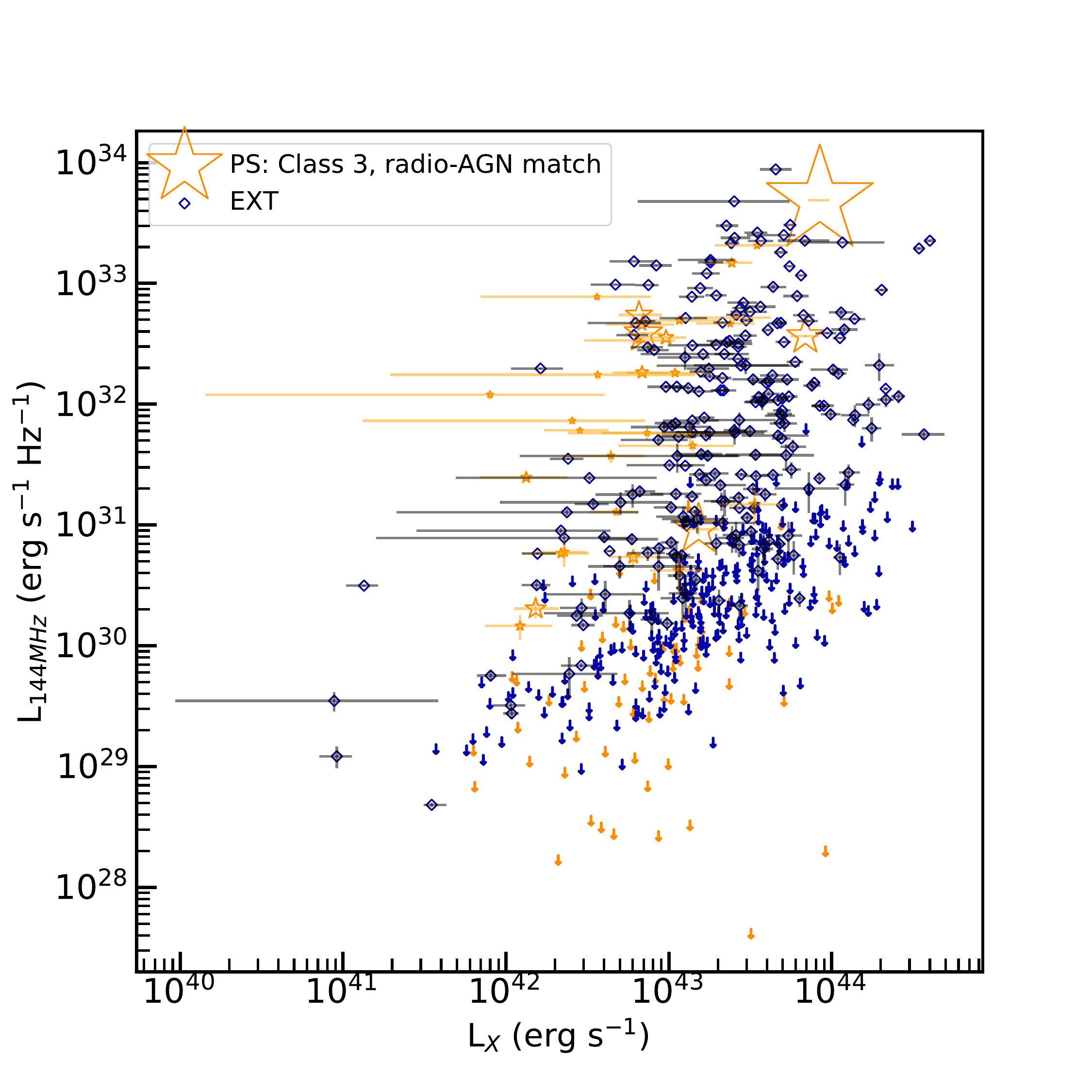}
\includegraphics[width=0.43\textwidth, trim=0 5 30 50, clip]{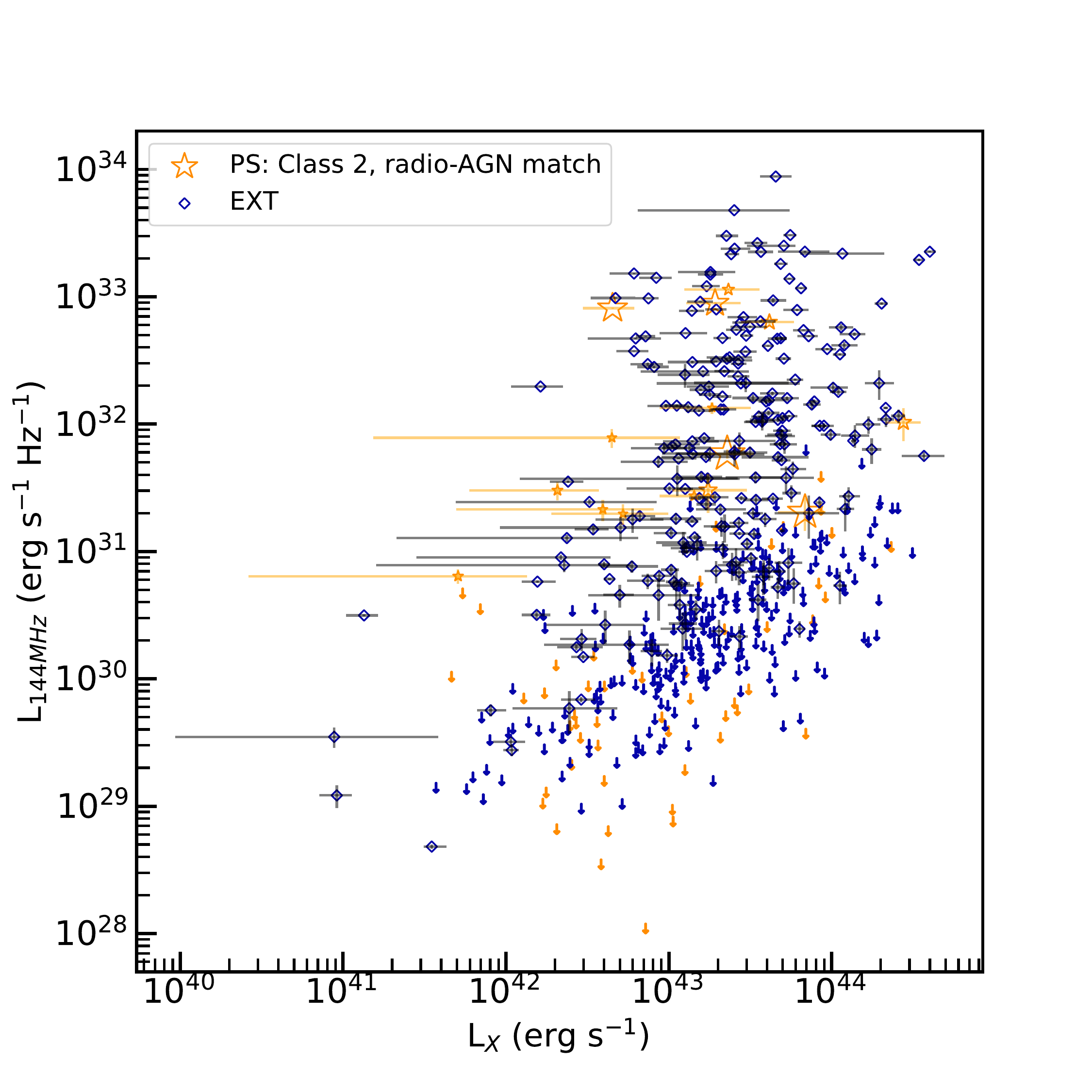}
\caption{Radio properties of the clusters in the point source sample compared with the extent selected sample. {\bf Top and middle panels}: Radio power of BCGs versus X-ray luminosity of the host cluster for the PS sample (orange) and for the cluster sample studied in P21 (blue). Data are sized based on the detection likelihood, and arrows denote radio upper limits. From topleft to bottom right: Clusters of class 5, 4, 3 and 2. {\bf Bottom panel}: Radio power measured at the position of the AGN vs. X-ray luminosity for Class 3 (left) and 2 (right) objects.}
\label{fig:lxlr}
\end{center}
\end{figure*}

Similarly to P21, we cross-matched radio sources to the BCG positions of each candidate cluster by setting a sky threshold $\sim 3\theta$, with $\theta$ being the synthesized beam of the interferometric radio observation. BCG positions were determined by selecting a massive red-sequence galaxy near the X-ray emission peak \citep[see][for further details]{Klein2021}. This matching method has a drawback in that we may not be able to associate the BCG with its radio emission in the case of very large sources if the radio coordinates lie farther away than 3$\theta$ from the galaxy. Nevertheless, using a looser threshold could produce false matches, and for this analysis, it is more convenient to lose some sources than to provide a large number of associations, some of which are insecure. 
After visual inspection of our results, we find 111 clusters that meet our criteria, while for 235 objects, we did not detect any radio source close to the BCG with the current sensitivity. Nondetections were treated as 3$\sigma$ radio upper limits, with $\sigma$ being the local \textit{rms} of the LOFAR observation estimated at the position of each cluster. Table \ref{tab:detnumbers} summarizes the composition of detections and upper limits. 

\begin{table*}
\caption{Number of radio detections and upper limits for the different classes using the two cross-matching methods. Results in common for the two methods are listed in the last row.}
\label{tab:detnumbers}
\centering
\large
\begin{tabular}{@{}llllllll@{}}
\toprule
Matching method &  & Class 2 & Class 3 & Class 4 & Class 5 & Total \\
N. of sources   & & 67 & 96 & 63 & 120 & 346\\
\midrule
\multicolumn{1}{l}{LOFAR-BCG} & \multicolumn{1}{l}{Detections} & 23 & 30 & 23 & 35 & 111  \\ 
\multicolumn{1}{l}{} & \multicolumn{1}{l}{Upper Limits} & 44 & 66 & 40 & 85 & 235 \\ \midrule
\multicolumn{1}{l}{LOFAR-AGN} & \multicolumn{1}{l}{Detections} & 15 & 33 & 27 & 17 & 92 \\ 
\multicolumn{1}{l}{} & \multicolumn{1}{l}{Upper Limits} & 52 & 63 & 36 & 103 & 254 \\ \midrule
 Detections in common &  & 10 & 20 & 23 & 14 & 67 \\ \bottomrule
\end{tabular}
\end{table*}

In Figure~\ref{fig:lxlr} we show the correlation between the radio power of BCGs and the X-ray luminosity of the host cluster for the different classes. The radio luminosity spans from $\sim5 \cdot 10^{28}$ to $\sim10^{34}$ W/Hz at 144 MHz. The range is the same as for the sample of BCG radio galaxies studied in P21. As also found in P21 and other similar works \citep[e.g.,][]{Pasini_2020, Pasini_2021b}, there is a trend for stronger radio galaxies to be hosted in more luminous (i.e., more massive) clusters. However, observational limitations, especially in the radio band, affect our results. This translates into a relevant number of upper limits. 
In P21, statistical tests were used to assess the correlation between X-ray and radio luminosities. Because each subsample includes relatively few objects, we cannot perform the same tests here. Instead, we focused on the differences among the different cluster classes with respect to the relation.

Class 4 represents the objects that are most securely classified as clusters. The point source counterpart from S21 is coincident with a red passive galaxy, which in most of the cases is also the BCG identified by MCMF. This class therefore shows a higher radio detection fraction ($\sim$36\%) than Classes 2, 3, and 5 ($\sim$34\%, $\sim$31\%, and $\sim$29\%, respectively), as expected for BCGs at the center of galaxy clusters \citep[e.g.,][]{Sabater2019}. On the other hand, other classes are expected to include AGN, either in projection or in their BCGs. This is reflected in their radio detection fraction: in Class 2, for example, we searched for radio emission at the position of the best counterpart from S21 that is likely not a BCG, or at least not at the center of a cluster. 

For this reason, we performed an alternative cross-matching. Instead of the BCG position from MCMF, we used the position of the {\sc NWAY} counterpart. We expect the number of matches of Class 4 clusters that are detected with this alternative method (LOFAR-AGN) to be similar to the first method (LOFAR-BCG) because the class is defined through the AGN and the cluster redshifts are the same, and the AGN is consistent with a red, passive galaxy. On the other hand, we should observe significant differences for Classes 2 and 3 because we now searched for radio emission at the position of an AGN that is different from the MCMF identified BCG. The results of this second cross-matching, performed using the same criteria as the first, are presented in Table \ref{tab:detnumbers}.
We find 27 matches for class 4, compared to 23 with LOFAR-BCG. All results of the first method are in common with LOFAR-AGN, indicating that the second method found only four more matches. The number of Class 5 radio detections drops from 35 (LOFAR-BCG) to 17 (LOFAR-AGN), as expected because for these objects, {\sc NWAY} was not able to locate any good counterpart.
The number of detections for Class 2 decreases (from 23 to 15), while for Class 3, it is similar (from 30 to 33). Overall, these numbers are not easy to interpret because they are driven by numerous factors such as the BCG and AGN position, the sensitivity of the radio observation at these positions, and the chosen matching criteria, combined with the fact that radio emission is not guaranteed even in the case of a secure cluster (see also P21), let alone when the identification as a cluster is not reliable. For example, if Class 2 objects are in reality AGN, we would expect that they do not follow the same trend as we observe for clusters in the relation of X-ray luminosity to radio power. A different correlation has been observed for AGN \citep{Vink2006, Sambruna1999}. To this end, we show in the lower panels of Figure~\ref{fig:lxlr} the distribution observed for Classes 2 and 3 with the second cross-matching method, LOFAR-AGN. Again, the low statistics does not allow us to derive conclusions from the correlation. Class 2 objects seem to stand out of the distribution observed for Class~4 clusters, suggesting that they might indeed represent different objects, but a larger sample is required to investigate this more accurately. In the near future, data from the \rosi\ all-sky survey (eRASS, Bulbul et al. in prep.) will provide a significantly larger sample of clusters and AGN, which,  combined with the LOFAR Two-Metre Sky Survey (LoTSS, \citealt{Shimwell2017}) as well as with the forthcoming LOFAR LBA Sky Survey (LoLSS, \citealt{deGasperin2021}), will help us to shed more light on this.

\subsection{Cross-matching the sample with the public SZ surveys}
\label{sec:crossmatch}
We cross-matched the clusters in the point source catalog with the public cluster catalogs based on observations of the SZ effect that cover the eFEDS field. If the cluster hosts a bright BCG dominated by the AGN, these massive clusters at higher redshifts could easily be confused as point sources due to their compact size and bright core emission by the X-ray telescopes with sizable PSFs, as in the case of the Phoenix and El-Gordo clusters \citep{McDonald2012, Menanteau2012}. Because of their redshift-independent sensitivity, SZ surveys are less prone to this effect provided that the flux and mass limits of the surveys are comparable to X-ray surveys. In this paper, we propose a method that locates the high-redshift cool-core clusters that are missed by the extent selection and the wide PSF of \rosi\ by carefully examining the point source catalog. To evaluate under which circumstances we miss high-redshift sources in the extent selection, we cross-matched our catalog with two independent public SZ surveys from the Planck and ACT telescopes that cover the eFEDS field. 

\begin{figure*}
\begin{center}
\includegraphics[width=0.95\textwidth, clip]{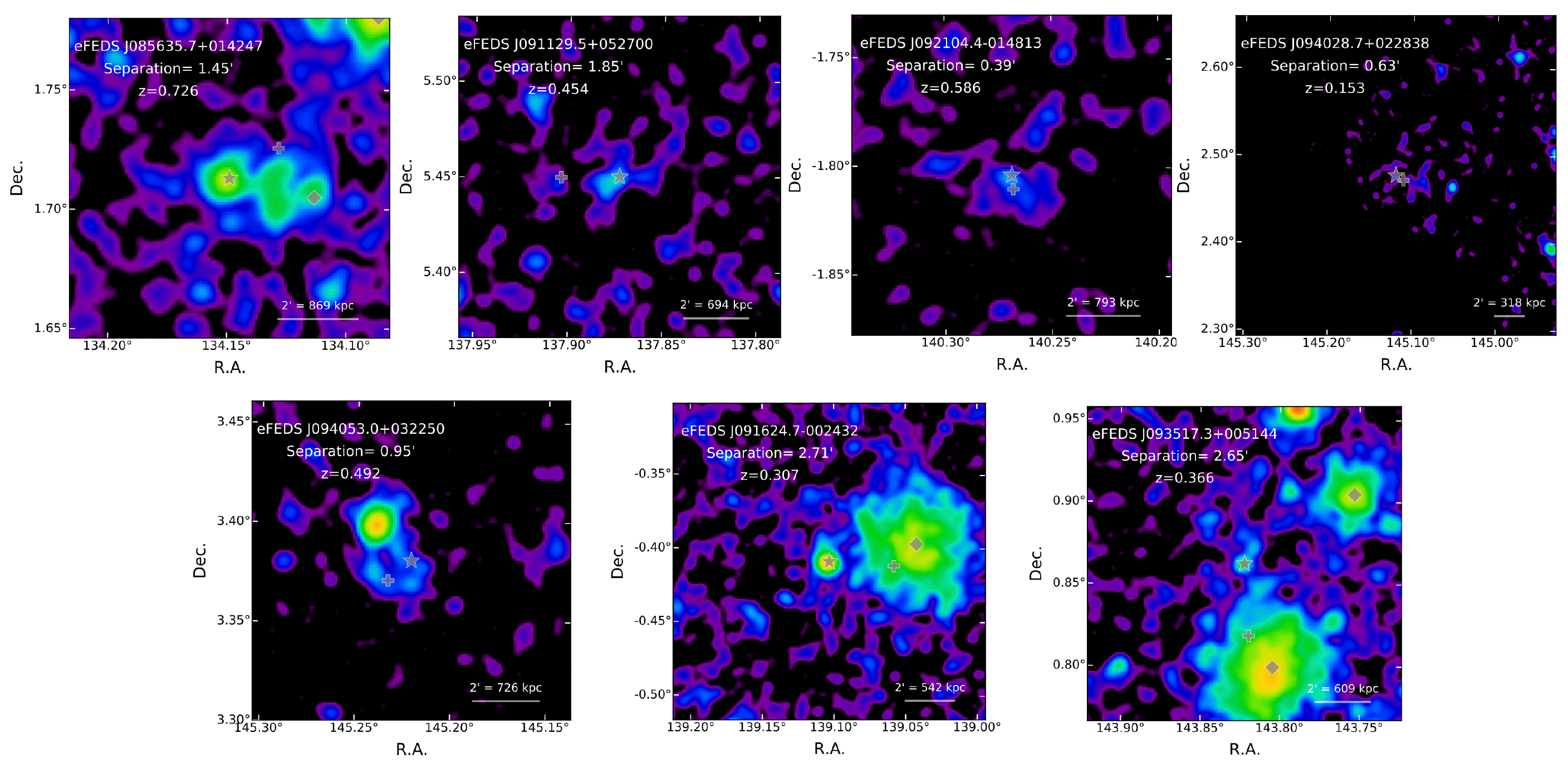}
\vspace{2mm}
\caption{Images of \rosi\ clusters identified in the point source catalog, matched with the public SZ catalogs \citep[ACT and Planck;][]{Hilton2021,2016planckb}.
{\bf Top panel:} Four matches (eFEDS~J084615.7+050708 at the edge of the field is ignored) with ACT using a matching radius of 2 arcmin. {\bf Bottom panel:} Three matches with PLANCKSZ2 using a matching radius of 3 arcmin. Images are centered on the X-ray position of the cluster in the point source catalog, marked by the gray star. The diamond shows the location of the extended sources in the \citet{Liu2021} catalog if there is any nearby. The SZ centroid from either the ACT or Planck catalog is shown as a cross.}
\label{fig:szcomp}
\end{center}
\end{figure*}

Specifically, we cross-matched our catalog with the SZ effect-based catalogs with ACT \citep {Hilton2021} and PLANCKSZ2 \citep{2016planckb}, which  cover the entire eFEDS field. 

Using a 2\arcmin\ search radius around the positions of the \rosi\ detected point sources, we find a total of five clusters that have counterparts in the ACT catalog. 
eFEDS~J085635.7+014247 (ACT CLJ0856.5+0143) is the highest-redshift cluster at $z=0.72$ in the matched sample. The cluster is marked as a Class~3 cluster in S21. The detection of this cluster in the point source catalog could be a case for a compact high-redshift cluster, and it would be misclassified as a point source by \rosi. The X-ray centroid in the point source catalog is located 1.45\arcmin\ southwest of the ACT centroid. Another cluster, eFEDS~J085627.2+014217, is also detected to the southeast of the ACT centroid in the extent-selected catalog at a very similar redshift \citep{Liu2021}. The cluster that is identified in this work by its red sequence and the cluster in the extent-selected sample probably belong to the same cluster system. Because we classify the cluster as a Class~3 object, that is, the point source identified here is in projection, this cluster in the point source catalog is either a merger and the point source in projection boosts the X-ray flux, or we detect the cluster outskirts in this case. In either case, the mass of this substructure or the subhalo is likely overestimated as a result of the point source flux. The ACT signal is likely to be associated with the more massive extended selected cluster with a mass of $M_{SZ, WL}=4.00^{+0.88}_{-0.77}\times10^{14}$~$M_{sun}$. The mass of the cluster eFEDS~J085627.2+01421 in the extended catalog is  6.58$^{+1.42}_{-1.12}\times10^{14}\ M_{sun}$ in \citet{chiu2021} and is fully consistent with the mass reported in \citet{Hilton2021}.

eFEDS~J091129.5+052700 (ACT CLJ0911.6+0526) at a redshift of 0.45, on the other hand, clearly hosts a bright core and could also be a merger as it shows an extended X-ray tail in the \rosi\ image. Given that the richness of this source is 41 and it is classified into Class~5 (without a secure point source counterpart), it is likely that we missed this compact low-mass halo marked with a star in Figure~\ref{fig:szcomp} because of the detection limit and extent likelihood cuts in the extent-selected sample. The ACT detection is located at 1.85$^{\prime}$ west of our detection and has a mass of $2.18^{+0.53}_{-0.43}\times10^{14}$~$M_{sun}$. Provided that the separation is greater than the beam size of ACT \citep[$1.4^{\prime}$ FWHM][]{Aiola2020}, the ACT and \rosi\ detections are likely to be low-mass clusters undergoing a merger activity. Our weak-lensing inferred mass of the halo detected in X-rays is $9.24_{-3.81}^{+5.46}\times10^{13}$~$M_{sun}$.

Similarly, the cluster eFEDS~J092104.4-014813 (ACT~CLJ0921.0-0148) at z=0.59 with richness 85 in Class~5 is detected by both ACT and \rosi. This is a clear case of a moderately massive galaxy cluster that was misclassified by our detection algorithm. The separation between the X-ray and the SZ centroid ($0.39^{\prime}$) is within the beam size of ACT and within the spatial resolution of \rosi. The reported mass of the cluster by the ACT collaboration is $M_{SZ,WL}$=$3.03^{+0.80}_{-0.67}\times 10^{14}$~$M_{sun}$, while we obtain a weak-lensing inferred mass of $M_{X,WL}=2.05_{-1.72}^{+1.09}\times 10^{14}$~$M_{sun}$. These measurements are consistent with each other within $1\sigma$ uncertainties. 

Another cross-matched cluster is eFEDS~J094053.0+032250 (ACT~CLJ0940.9+0322) at a redshift of 0.49,  which is  classified as a Class~4 cluster with richness 37. This cluster is located $1.5^{\prime}$ away from a very bright point source eFEDS~J094057.2+032359 at z=0.06 in S21. Because this is a lower-mass system, the influence from the very bright point source nearby might be another reason for the misclassification of this cluster. The \rosi\ reported mass $M_{X,WL}=2.41_{-0.76}^{+0.89} \times10^{14}$ $M_{sun}$ and the ACT reported mass of 2.76$^{+0.77}_{-0.63}\times10^{14}$ $M_{sun}$ agree. The photo-z redshift measurements of these clusters are consistent with each other within 1\% in the two catalogs. The agreement of SZ and X-ray masses indicates that our X-ray postprocessing is able to remove the contamination from the northern point source well, but the large separation between the centroids ($0.95^{\prime}$) could be due to the bias caused by the bright point source emission in X-rays and SZ during the detection run.  One property in common for all these clusters is that they all have a low SZ signal-to-noise ratio of $\sim5$ in the ACT catalog, suggesting that \rosi\ probes the low-mass and low-luminosity cluster population with a low signal-to-noise ratio that is detected by SZ telescopes, as is the case of the extent-selected samples. 

The last matched cluster is eFEDS~J084615.7+050708 (ACT-CL J0846.2+0508). However, this cluster is located at the edge of the eFEDS field. We therefore do not discuss this case in detail. 

We also cross-matched our sample with the PLANCKSZ2 catalog using a matching criterion of 2\arcmin\ because the beam size of the Planck Telescope is larger \citep{2016planckb}. We find one cluster, eFEDS~J094028.7+022838 (PSZ 2G232.84+38.13), at redshift $z=0.15$, that is detected by both \rosi\ and Planck. The Planck detection is 38$^{\prime\prime}$ away from the \rosi\ X-ray center in the point source catalog. S21 flags this cluster as a Class~2 object because there are no nearby extent-selected clusters. This source is likely to be an AGN-dominated low-mass cluster with a richness of 73. However, this source is located at the border of the eFEDS field, which makes the luminosity and mass measurements challenging from the X-ray data. The \rosi\ mass $ 2.18_{-0.45}^{+0.43}\times 10^{14}\ M_{sun}$ of this cluster still agrees with the Planck measurement of $3.43\pm 0.3\times10^{14}\ M_{sun}$ at about 2$\sigma$. Because of the large beam size of Planck, we extended our search criteria to 3\arcmin\, and located two other clusters that were cross-matched with the eFEDS point source cluster sample. PSZ2~G232.79+31.48 is matched with eFEDS~J091624.7-002432 (Class~3) in the point source catalog with a centroid separation of $2.71^{\arcmin}$ between Planck and \rosi. We also find an extent-selected cluster nearby, eFEDS~J091610.2-002348 in \citet{Liu2021}, located 3.4\arcmin\ away to the east. The Planck detection is a better match to the extent-selected cluster eFEDS~J091610.2-002348 considering the smaller separation of centroids. eFEDS~J091610.2-002348 (PSZ2~G232.79+31.48) is detected in the extent-selected sample with a mass of 5.4$_{-1.3}^{+2.8}\times10^{14}$~$M_{sun}$ and is consistent with the Planck-reported mass of (4.86$\pm$0.63$\times10^{14}$~$M_{sun}$). In this case, either our MCMF algorithm picked up the red galaxy distribution from the outskirts of the extent-selected cluster eFEDS~J091610.2-002348, or we detected an infalling low-mass group onto the more massive cluster. As expected because of their small separation together with the large Planck beam size, individual clusters were not resolved by Planck. \rosi,\, on the other hand, was able to resolve each, demonstrating the power of X-ray surveys in mapping large structures and locating a wide population of clusters of galaxies. Another similar case is eFEDS~J093517.3+005144 (Class~3) at a redshift of 0.37 and richness of 98, which is matched with the nearby Planck cluster PSZ2~G233.68+36.14, but the SZ signal is 2.65$^{\prime}$ away from the centroid of the X-ray source. We find two nearby clusters, one in the north (eFEDS J093500.7+005417) and the other in the south (eFEDS J093513.0+004757) in a pair at the same redshift of $\sim0.35$ in the extent-selected catalog in close proximity to this source. The Plank detection is a better match to the southern massive cluster. Our detection in the point source catalog could either be the outskirts of the massive southern cluster or a group infalling onto eFEDS J093513.0+004757 considering their very similar redshifts and proximity. 

Although the beam size of Planck is large, the 3$^{\prime}$ matching radius is already too large for cross-matches of the \rosi\ and other SZ catalogs. It is important to note that all of the matched clusters presented here have a low signal-to-noise ratio $<$6 and low mass in both ACT and Planck catalogs. This shows that \rosi detects rather low-mass systems that are just above the detection limit of the current SZ experiments.

\section{Clusters hosting active galactic nuclei}
\label{sec:agnclusters}

Another goal of this study is to identify clusters and groups that host bright cool-cores and AGNs. In cooling-core clusters, bright X-ray emission and mid-IR emission is provided by the warm dust heated by the AGN or star formation, and radio emission from AGN jets is often observed together. Accretion of a central black hole that is fueled by the cooling gas can also produce radio emission in strong cooling-core clusters. Radio jets and lobes associated with the AGN may affect the surrounding ICM gas, creating deficits in the X-ray emission around the radio jets, which are called ``cavities"  in cluster centers \citep{Blanton2010, Hlavacek-Larrondo2012}. Several clusters show a clear association of a cooling flow in a cluster with the central galaxy (and AGN) that hosts radio sources. Examples are Abell~2052 \citep{Blanton2001}, Abell~2597 \citep{McNamara2001}, Abell~4059 \citep{Heinz2002}, Abell~262 \citep{Blanton2004}, and MS~0735.6+7321 \citep{McNamara2005}. Because of their compact and peaked X-ray emission, these clusters would be misclassified as point sources by the detection algorithm.

Our goal is to find these cool-core clusters with AGN in their BCGs using the multiwavelength data through \rosi, LOFAR, WISE, and SDSS in the eFEDS point source catalog. Therefore, we searched for clusters that are bright in X-rays, radio, and mid-IR relative to the optical data in the eFEDS field. The catalog of clusters presented in this work was constructed by running the MCMF around the X-ray centroids. In general, MCMF and similar red-sequence galaxy concentration-finder algorithms such as redmapper \citep{Rykoff2014} are based on finding red galaxies and generally identify a red galaxy as an optical counterpart. The BCG identified by MCMF is often a red galaxy because of the nature of the algorithm and is unlikely to be an AGN-hosting BCG. Therefore, MCMF would miss blue BCGs that might host a bright quasar at the center. On the other hand, our other algorithm {\sc NWAY,}  which was used to identify the counterparts of the point-like sources in S21, searches for all possible counterparts with a broader prior than only red-sequence features and would locate both the AGN and red galaxies as potential counterparts around the X-ray centroid. Therefore, a combination of the two methods, that is, using MCMF to find secure clusters through their red sequence with a  probability $f_{cont}>0.2$ and matching those with counterparts found through {\sc NWAY} is the most powerful way to find potential BCG candidates that might host AGN. We then examined their optical, IR, and X-ray properties to construct a sample of BCG candidates. We summarize our procedure below to identify these special cases.

\begin{enumerate}
\item We constructed a secure cluster sample through red-sequence galaxy concentration in close proximity to an X-ray point source with MCMF. 

\item We categorized the full sample with {\sc NWAY} based on the redshift and proximity of the point source counterpart to construct cluster classes.

\item We built a subsample of clusters that host a point source at the cluster redshift near the X-ray and optical centroids. These are the Class~3 clusters in the S21 classification scheme.

\item We applied a color selection cut to identify clusters that are outliers relative to the rest of the cluster population and bright in the IR band.

\item We examined the optical spectra and the X-ray, IR, and radio emission associated with the host cluster in this subsample. 
\end{enumerate}

In this section, we describe the color-color diagrams, X-ray, radio, IR, and optical properties of these clusters. We note that we did not attempt to detect the cavities around the radio sources due to the PSF of \rosi\ and the depth of the survey. Follow-up \chandra\ or \xmm\ observations are required for detailed imaging observation measurements.

\begin{figure*}
\begin{center}
\includegraphics[width=0.49\textwidth]{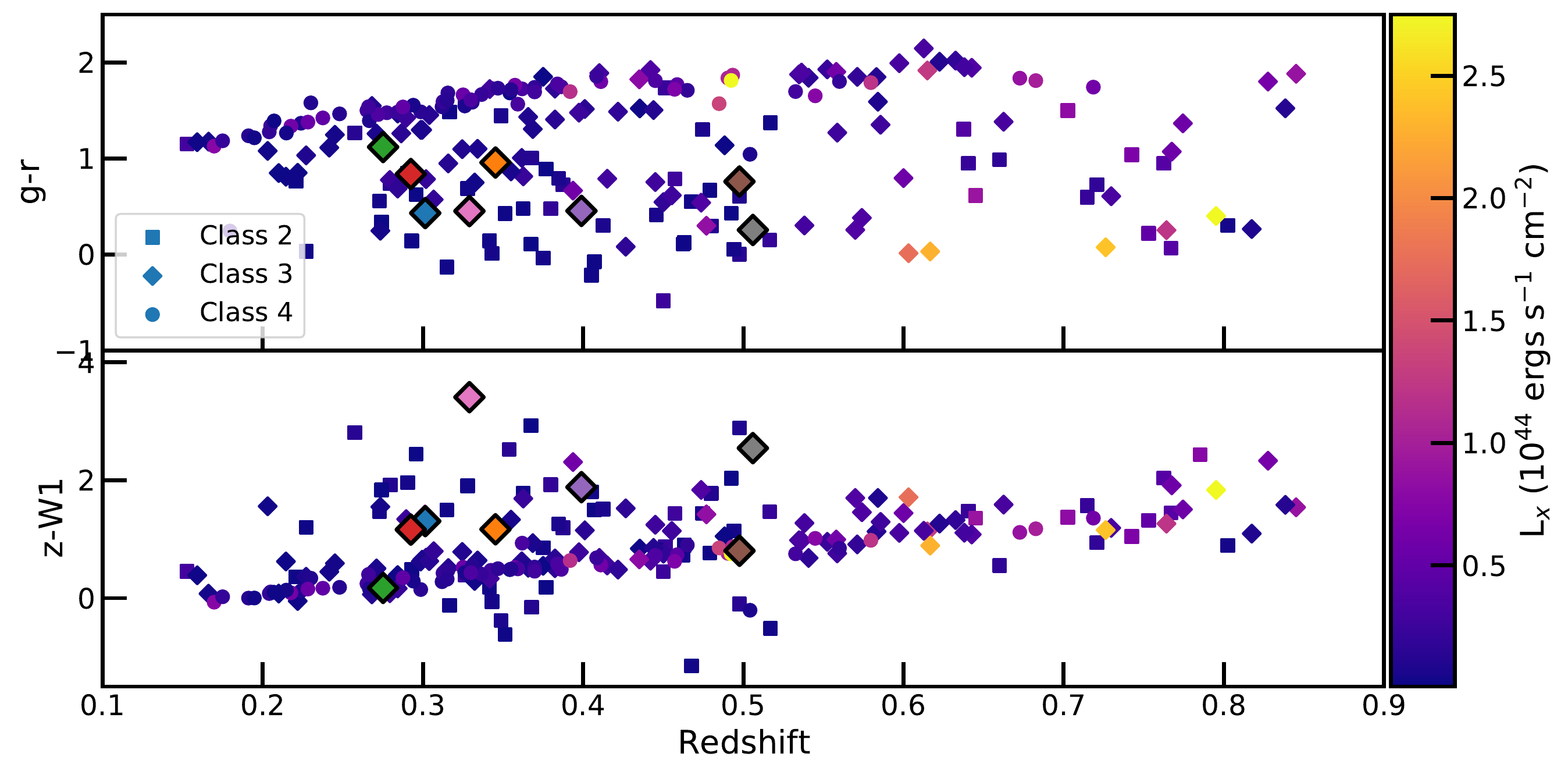}
\includegraphics[width=0.49\textwidth]{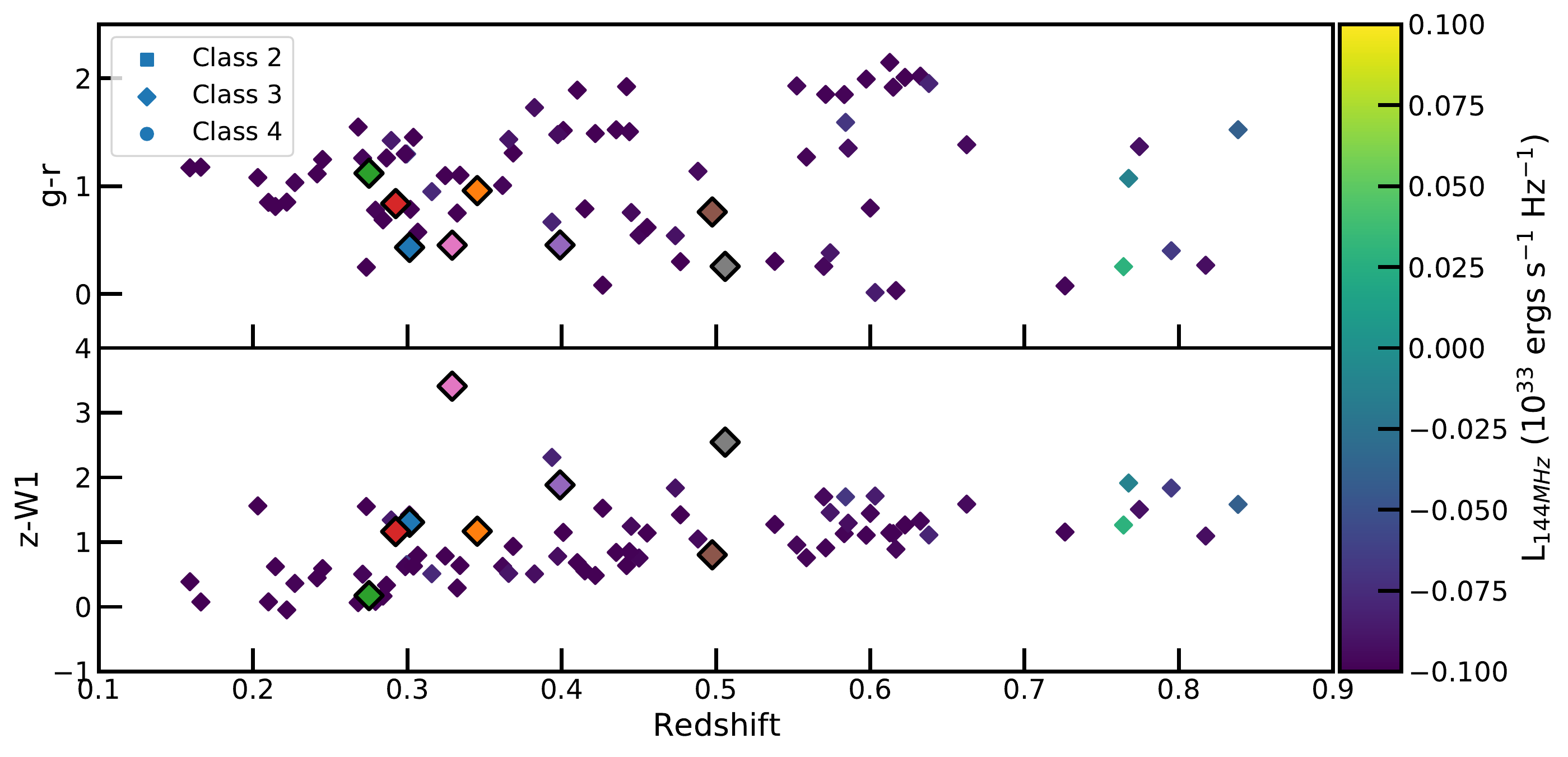}
\includegraphics[width=0.49\textwidth]{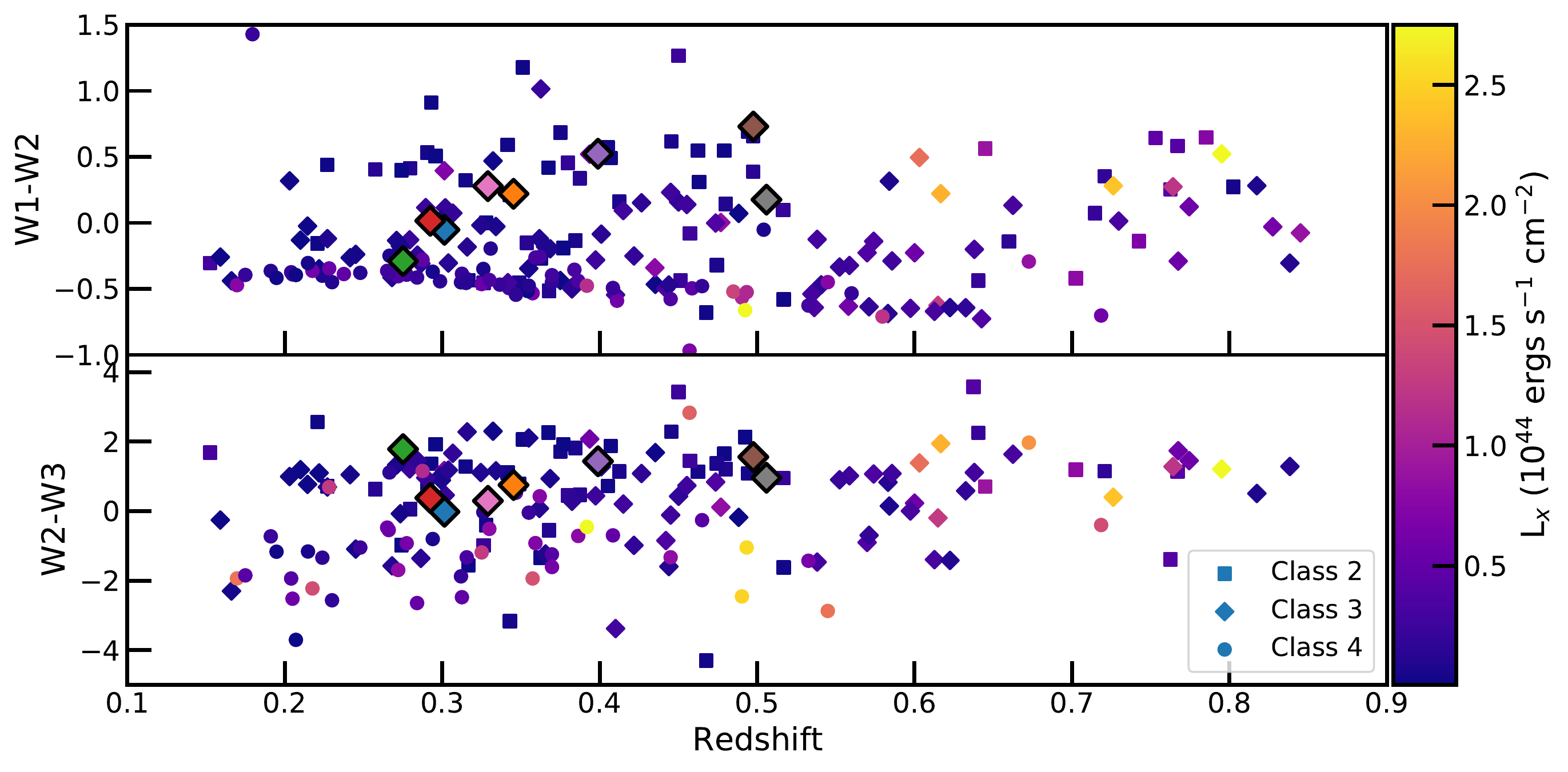}
\includegraphics[width=0.49\textwidth]{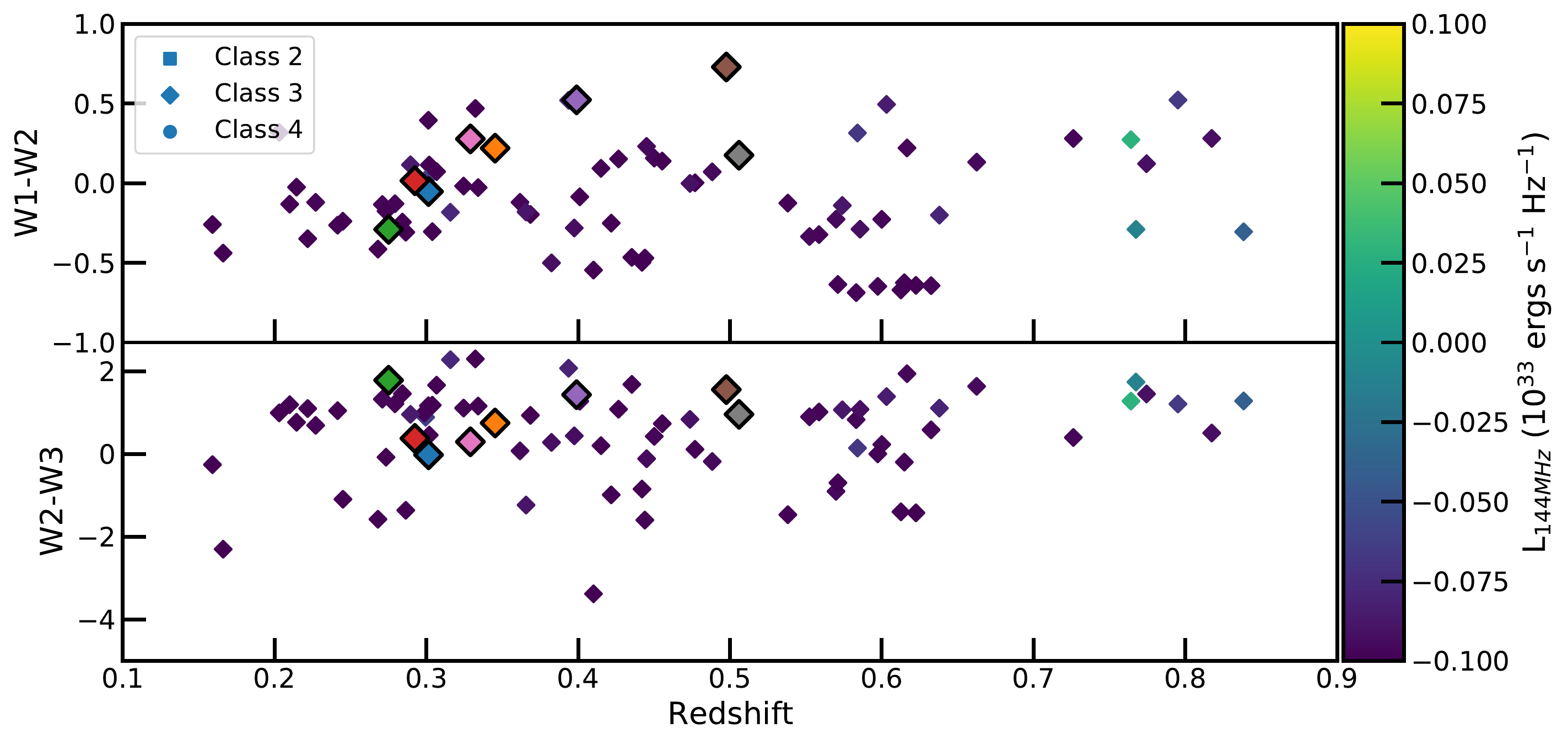}
\caption{Color redshift diagrams of the Class~2, 3, and 4 clusters. {\it g-r, z-W1, W2-W3, \textup{and} W1-W2} color diagrams of the counterparts identified by {\sc NWAY} as a function of X-ray (panels on the left) and radio (panels on the right) luminosities of the host clusters in Classes 2, 3, and 4 with secure {\sc NWAY} and MCMF counterparts. The outlier clusters are shown in black. They include the Class~3 clusters and groups with radio loud-AGN in their BCGs selected based on our color criteria. All these four cases have SDSS spectroscopy for further investigation.}
\label{fig:colordiagrams}
\end{center}
\end{figure*}
\subsection{Selection based on color-redshift diagrams}
\label{sec:color}
One way to identify the AGN-dominated BCGs is through their colors in optical and IR bands. Specifically, we searched for clusters that deviate from the bulk of the cluster population in color-color diagrams. AGN can heat a larger fraction of their surrounding dust to higher temperatures and cause a relatively stronger emission at MIR wavelengths with respect to the apparent flux deficit at far-infrared (FIR) wavelengths \citep{Mullaney2011, lee2013}. Moreover, the AGN activity should trigger the mid-IR (MIR) luminosities that are sensitive to star formation and AGN activity. Radio and infrared fluxes are also good diagnostics for locating the AGN that is ``active". \citet{Green2016} find that AGN-hosting BCGs appear redder than the passive populations in the $W2-W3$ and $W1-W2$ bands, whereas the star-forming BCGs appear red only in $W2-W3$ \citep{Green2016}. We expect that BCGs that host AGN are redder than the passive galaxies, and this would be reflected in the $g$ and $r$ and WISE $W1$ and $W2$ colors of the source. In Figure~\ref{fig:colordiagrams} we present the color-redshift diagrams in bands $z-W1$, $g-r$, $W1-W2$, and $W1-W3$ of the sample clusters and their {\sc NWAY-} identified counterparts of the clusters in Class~2, 3, and 4. Class~5 clusters were removed from the figure as they lack secure {\sc NWAY} counterparts. The side color bars demonstrate the observed X-ray and radio luminosities of these clusters within R$_{500}$. Here we use the optical and infrared colors presented in S21.

We started with a sample of potential clusters and AGN at the same redshifts based on their class, that is, Class~3 clusters (112 clusters in total). We then examined their color-redshift distributions in the $g-r$, $z-W1$,  $W1-W2$, and $W2-W3$ bands to identify clusters that are redder than the population of red galaxy members. In the redshift distributions of the $g-r$, $W1-W2$ , and $z-W1$ in Figure \ref{fig:colordiagrams}, we see that the population of red galaxies mostly consists of Class~4 clusters, whose counterpart red galaxy is identified in {\sc NWAY} and MCMF. We do not see any particular correlation in the $W2-W3$ bands, therefore we excluded these bands from further selection. A subsample of 46 clusters shows significant color offsets from the rest of the population with the selection criteria of $g - r <1.15$, $W1 - W2>-0.3$, and $z-W1>$-0.5. These outlier clusters are all likely to host a radio-quiet or radio-loud AGN and would be interesting targets to examine further. We examine the clusters with central radio-loud AGN in the next section.

\subsection{Clusters bright in X-rays, IR, and radio}
\label{sec:color2}

\begin{figure*}
\begin{center}
\includegraphics[width=1.\textwidth]{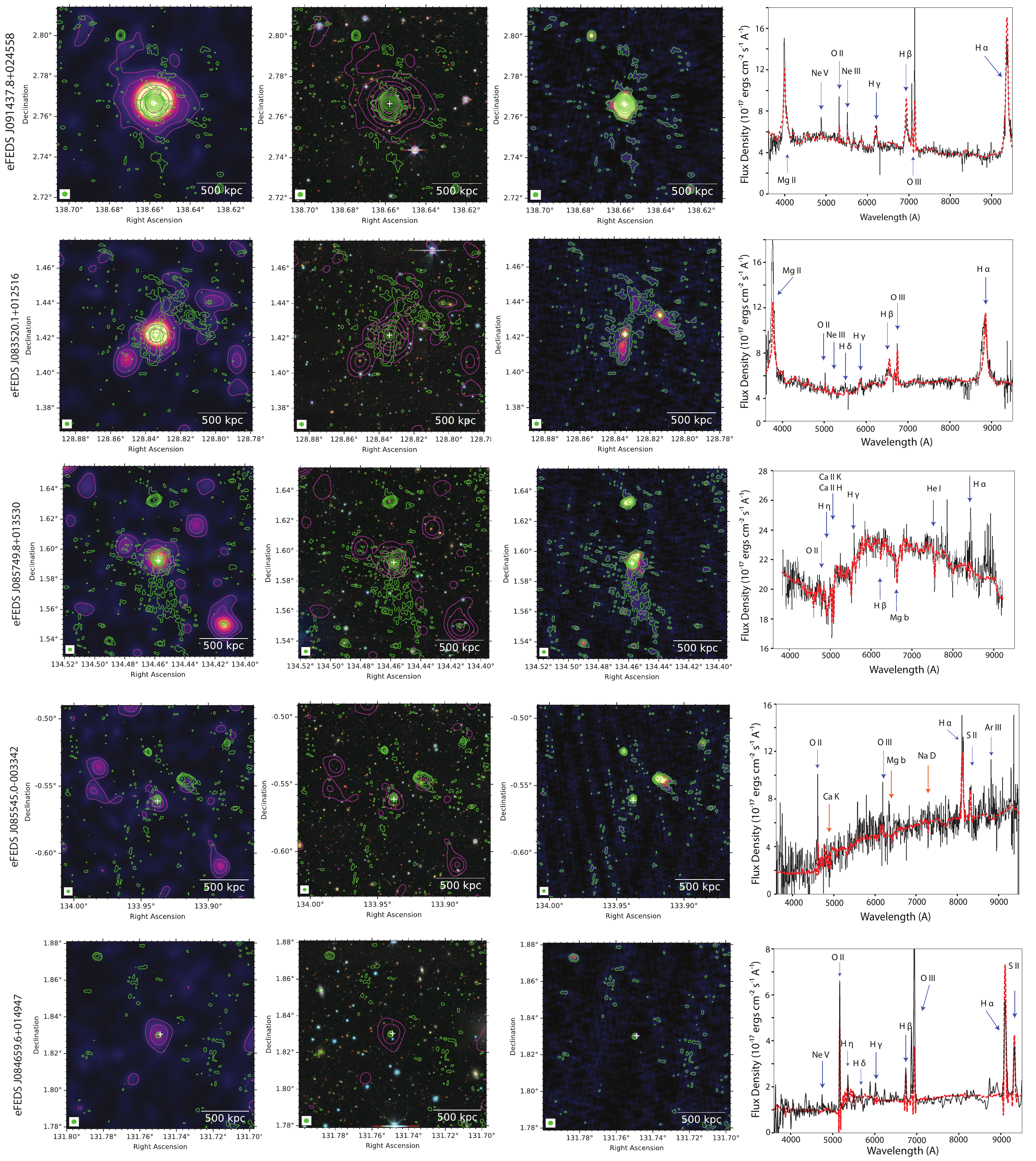}
\caption{$0.5-2$~keV soft-band images of the AGN-hosting clusters and groups in Class~3. The soft-band eROSITA images are shown in the leftmost panel. In the two middle panels, we provide DECaLS images in the optical band and LOFAR images in the radio band of the same clusters. The red-sequence galaxies are shown in the DECALS images. The overlaid LOFAR contours in green show the location of the radio source associated with the identified BCGs from {\sc NWAY}. The rightmost panels show the SDSS spectra of the counterpart identified by the {\sc NWAY}. Green lines show the significantly detected emission lines, whereas orange lines show the absorption lines on the optical SDSS spectra displayed in the right panels. The continuous line in red shows the best-fit model to the spectra.}
\label{fig:selected}
\end{center}    
 \end{figure*}

\begin{figure*}
\begin{center}
\includegraphics[width=0.95\textwidth]{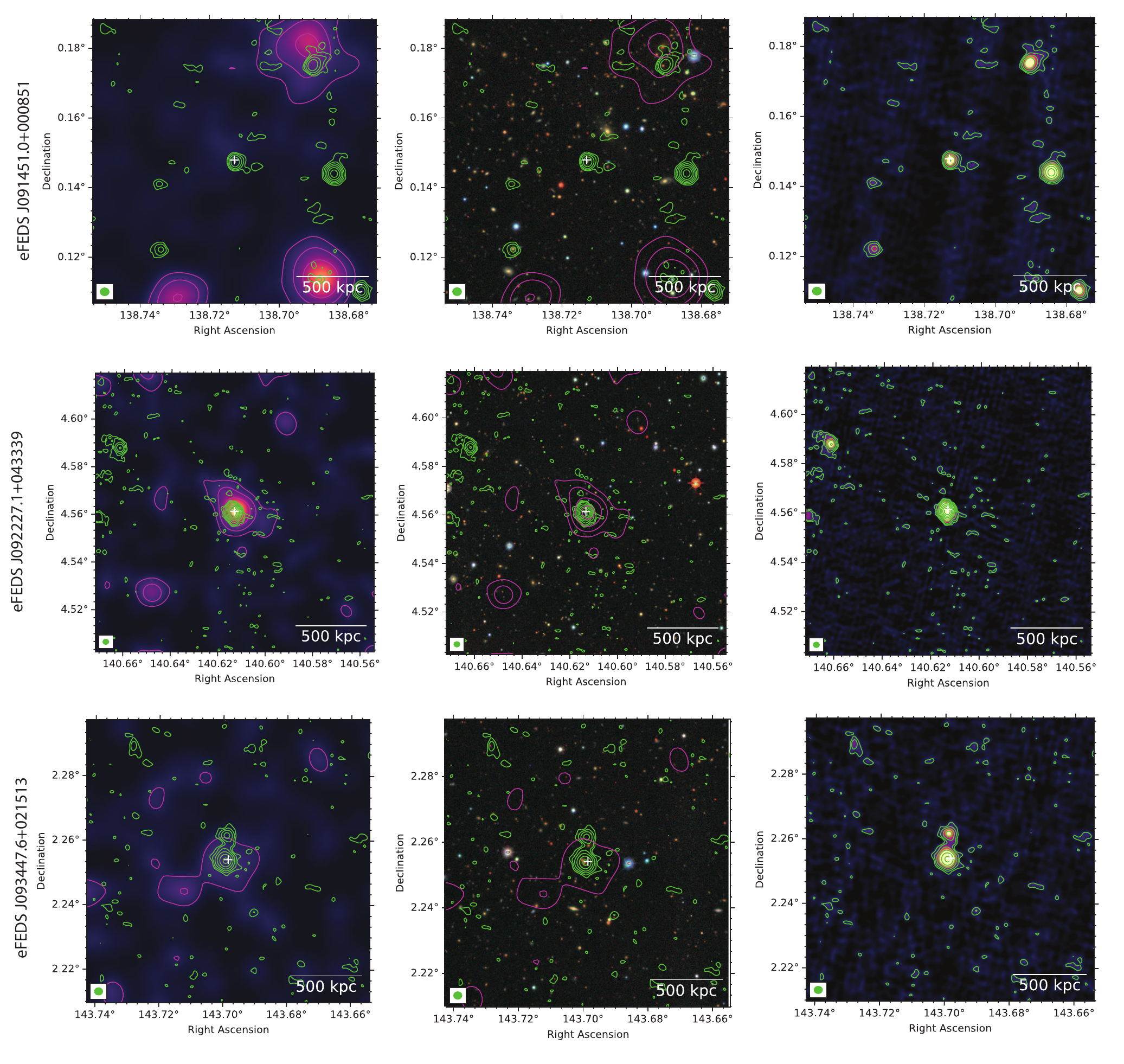}
\caption{$0.5-2$~keV soft-band images of the candidates of clusters and groups hosting radio-loud AGNs in their BCG in Class~3. The soft-band eROSITA images are shown in the leftmost panel. The two middle panels show DECaLS images in the optical band and LOFAR images in the radio band of the same clusters. The red-sequence galaxies are shown in the DECALS images. The overlaid LOFAR contours in green show the location of the radio source associated with the cluster. }
\label{fig:selected_nosdss}
\end{center}    
 \end{figure*}

To construct a sample of blue BCG candidates, we searched for clusters with significant, radio, X-ray, and IR emission while hosting an AGN in their vicinity at the same redshift, and show significant color offset from the rest of the red sequence galaxies. Out of these 46 clusters in the Class~3 group, we find that 33 of them have a LOFAR radio detection within their R$_{500}$ and show significant color offset in the $g-r$, $z-W1$, and $W1-W2$ planes. We identified a subsample of eight BCG candidates with a significant LOFAR radio detection ($>5\sigma$ significance) in close proximity to the X-ray detection, within the positional accuracy of eROSITA \citep[$\sim 5^{\prime\prime}$, see][]{Brunner2021}. These candidates are shown in the color-color diagrams in Figure~\ref{fig:colordiagrams} and are marked with diamonds with black corners. Only five of these eight clusters have SDSS spectra (see Merloni et al. 2022 for spectral targeting). We describe their multiwavelength properties through \rosi\ X-ray, LOFAR radio, DECALS optical data, and SDSS-V spectra below and provide their multiwavelength images in Figure~\ref{fig:selected}. The SDSS-IV spectra of the five counterparts show clear signatures of AGN and absorption lines from the host galaxy. In five cases, we detect emission lines from the AGN and absorption lines in their SDSS spectra. This confirms our classification scheme. The remaining three clusters were not targeted with SDSS, but we still provide their images in the X-ray, optical, and radio bandpasses in Figure~\ref{fig:selected_nosdss}.

\begin{enumerate}

    \item {\bf eFEDS~J091437.8+024558} is the most spectacular cluster identified in this subsample with a mass of 3.61$_{-0.36}^{+0.47}\times10^{14}$ $M_{sun}$ at z=0.4. Significant radio emission is associated with the central AGN, as shown in Figure~\ref{fig:selected}. The X-ray and radio luminosities are  8.42$_{-1.24}^{+1.40}\times10^{43}$~erg~s$^{-1}$ and 0.913$\pm$0.003~Jy. The colors of the counterpart show IR excess emission with $g-r=0.45$, $z-w1=1.88$, and $w1-w2=0.52$. The bright radio source is detected at 1.6$^{\prime\prime}$ away from the X-ray detection. The optical spectrum shows significantly bright and broad H$\alpha$ and $\beta$ lines from the central AGN.
\vspace{2mm}

    \item {\bf eFEDS J083520.1+012516} is a low-mass cluster located at a redshift of z=0.30 with a mass of $1.15_{-0.18}^{+0.25}\times10^{14}$~$M_{sun}$. The X-ray luminosity inferred from the scaling relations is $ 1.52_{-0.30}^{+0.47}\times10^{43}$~erg~s$^{-1}$ and the radio luminosity is $0.47\pm 0.05\times 10^{-2}$~Jy. Very extended AGN radio emission with an elongated filament ($\sim$800 kpc) to the northwest of the system is shown in Figure\ref{fig:selected}. The colors of the counterpart, $g-r=0.43$, $z-w1=1.31$, and $w1-w2=-0.05$, show excess in the the infrared bands. The SDSS spectra of this system show broad H$\alpha$ lines from the central QSO and Ca-H and Ca-K emission lines from the host galaxy.
   \vspace{2mm}

   \item {\bf eFEDS J085749.8+013530} is most likely a group with richness 26 at a redshift of 0.29. The soft-band X-ray luminosity is $6.97_{-1.54}^{+2.04}\times10^{42}$ ergs s$^{-1}$ and the mass is $6.46_{-1.04}^{+1.38}\pm 10^{13}$ M$_{sun}$. The colors also show departures from the main sequence with $g-r=0.84$, $z-w1=1.16$, and $w1-w2=0.02$, The radio luminosity of the central radio source is $ 0.14\pm 0.002$~Jy. The AGN emission in the center with filaments of radio emission extends to the south for $\sim$500 kpc. This source was also targeted with the SDSS-IV. The spectrum of the counterpart  shows a clear sign of H$\alpha$ emission from the central AGN, as well as absorption features from the host galaxy.  \\
\vspace{2mm}

    \item {\bf eFEDS J085545.0-003342} is a clear case of a nearby group with a radio-loud central AGN. The red galaxy BCG identified by MCMF and the AGN identified by {\sc NWAY} are located at the same redshift, z=0.28. The total mass inferred from the scaling relations is $1.95_{-0.69}^{+1.11}\times10^{13}$ $M_{sun}$. The soft-band X-ray luminosity is $1.33_{-0.64}^{+1.05}  \times10^{42}$~erg~s$^{-1}$. The colors show mild excess in the infrared with $g-r=1.12$, $z-w1=0.17$, and $w1-w2=-0.29$. We kept it in the list for completeness. We also find significant radio emission associated with the AGN with a LOFAR luminosity of $5.81\pm 0.15 \times10^{-2}$~Jy. The optical spectrum shows clear absorption lines from the host galaxy, as well as a weak broad H$\alpha$ line from the obscured AGN.
     
\vspace{2mm}
   \item {\bf eFEDS J084659.6+014947} is another low-mass group with a richness of 19 at a redshift of 0.35. The soft-band X-ray luminosity is $ 6.04_{-1.75}^{+2.98} \times10^{42}$ ergs s$^{-1}$. The total mass is estimated to be $5.66_{-1.25}^{+1.92}\times 10^{13}$ M$_{sun}$. The colors show mild infrared excess with $g-r=0.96$, $z-w1=1.17$, and $w1-w2=0.22$. The source has a weak radio source with a luminosity of $0.14\pm 0.02$~Jy at the location of the X-ray detection. The optical spectra show a bright H$\alpha$ line from the central AGN. 

\vspace{2mm}
    \item {\bf eFEDS J091451.0+000851} is a high-redshift (z=0.5) low-mass group with a richness of 29. The soft-band X-ray luminosity is $2.55_{-2.42}^{+4.59}\times10^{42}$ ergs s$^{-1}$, and the estimated mass is $3.17_{-2.79}^{+3.37}\times 10^{13}$ M$_{sun}$. The counterpart has colors of $g-r=0.76$, $z-w1=0.80$, and $w1-w2=0.73$ with brighter emission in the infrared than the red member galaxies. The radio source with a luminosity of $0.82\pm 0.04 \times10^{-2}$~Jy is significantly detected at the $>22\sigma$ confidence level. This source is not observed by SDSS-IV, therefore it is not possible to confirm the classification of this source.

\vspace{2mm}

   \item {\bf eFEDS J092227.1+043339} is another low-mass group with a richness of 15 at a redshift of 0.33. The  soft-band X-ray luminosity is $6.55_{-1.63}^{+2.53}\times10^{42}$ ergs s$^{-1}$, and the estimate total mass inferred from the eFEDS scaling relations is $ 6.55_{-1.63}^{+2.53}\times 10^{13}$ M$_{sun}$. The counterpart shows the brightest IR emission with colors $g-r=0.45$, $z-w1=3.41$, and $w1-w2==0.28$. The bright radio source detected 1.7$^{\prime\prime}$ away from the X-ray detection is significantly detected at a luminosity of $ 0.16\pm 0.002$~Jy, but this source lacks optical spectroscopy. 
    
\vspace{2mm}
     \item {\bf eFEDS J093447.6+021513} is another low-mass group at a redshift of 0.50. The soft-band luminosity is $6.49_{-3.47}^{+4.41} \times10^{43}$~erg~s$^{-1}$. The group has an estimated total mass of $ 5.43_{-2.32}^{+2.58}\times10^{13}$ $M_{sun}$. The colors show bright IR emission with $g-r=0.26$, $z-w1=2.54$, and $w1-w2=0.18$. The LOFAR-detected radio source located  $5^{\prime\prime}$ away from the X-ray detection has a luminosity of $ 3.63\pm 0.08\times 10^{-2}$~Jy. No SDSS spectroscopy is available for this source.

\end{enumerate}

In our visual inspection, we identified several other promising candidates, but they are not as strong radio and X-ray emitters as the candidates above. They include eFEDS~J091218.7+022924, eFEDS~J093833.0+025017, eFEDS~J085902.4-005150, and eFEDS~J091649.2+000033, which show an AGN with clearly visible Ca-H and Ca-K and broad H$\alpha$ lines, as well as emission from the red galaxy in their optical spectra.

\begin{figure*}
  \includegraphics[width=0.32\textwidth]{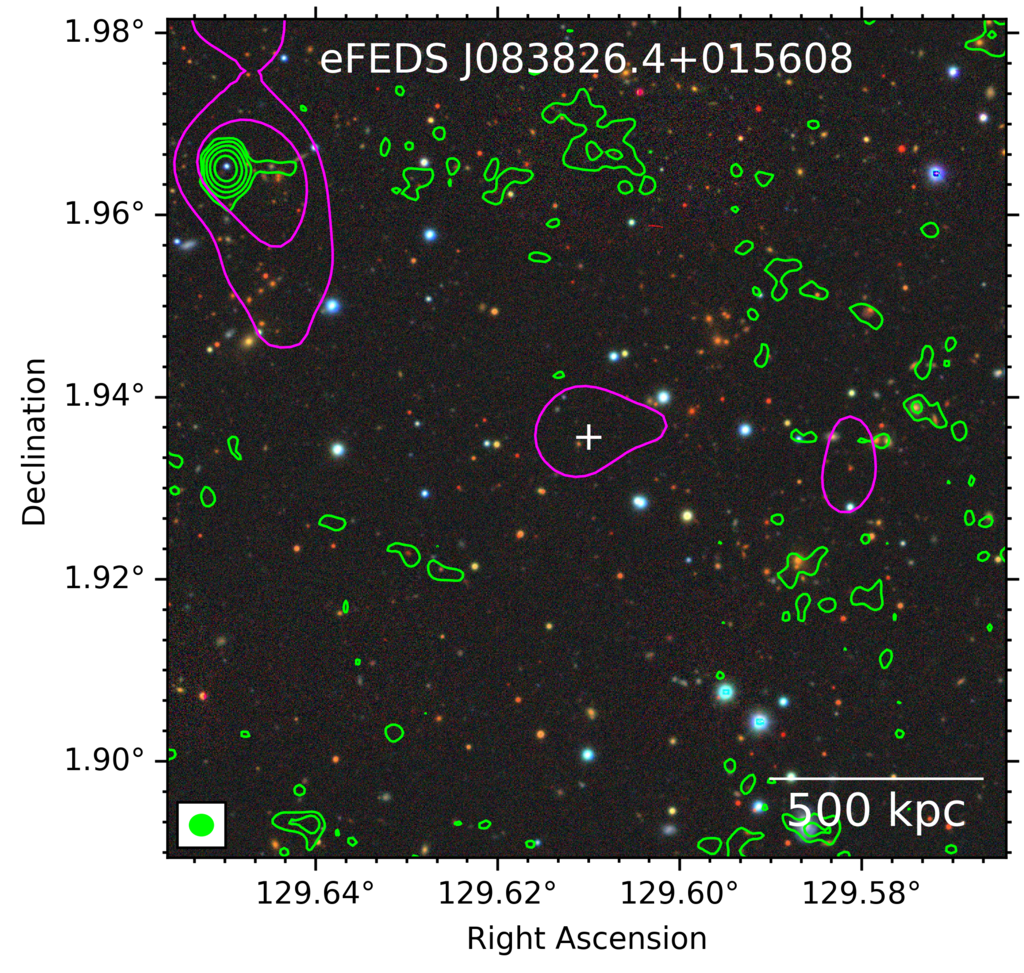}\hfill
  \includegraphics[width=0.32\textwidth]{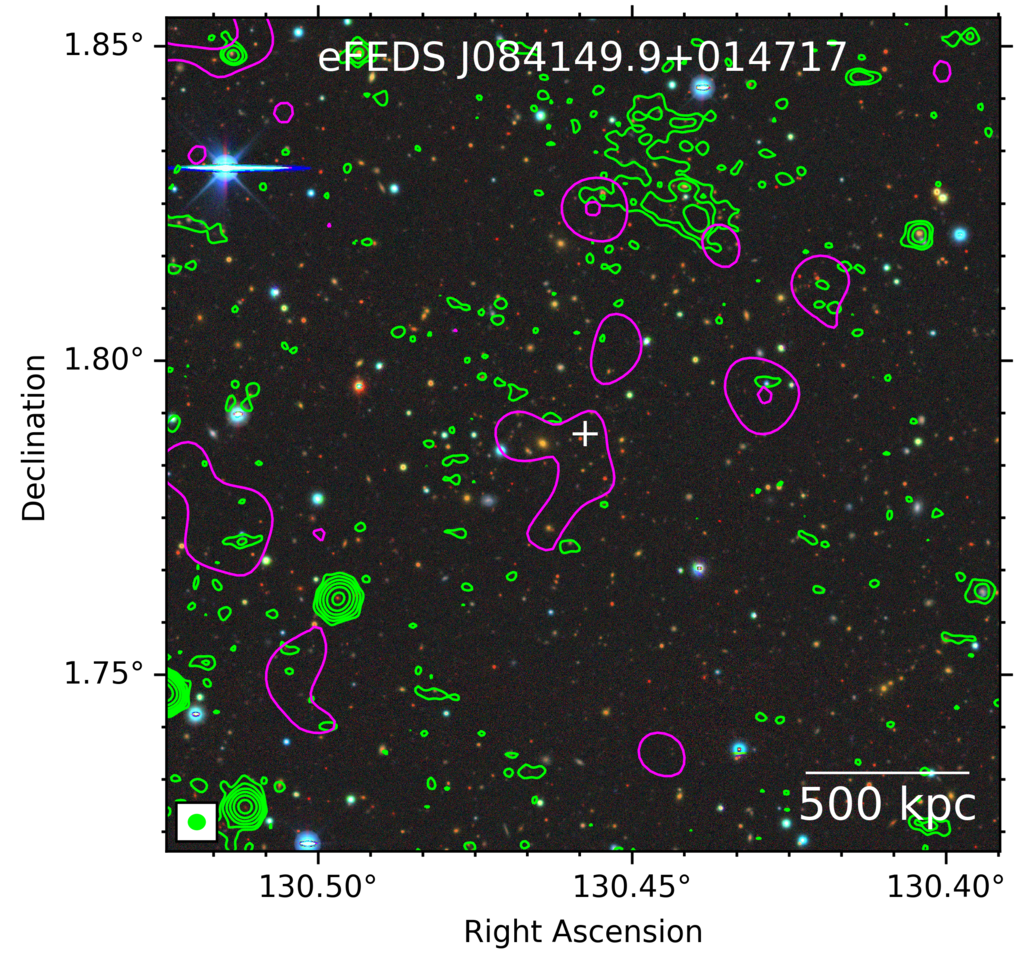}\hfill
  \includegraphics[width=0.32\textwidth]{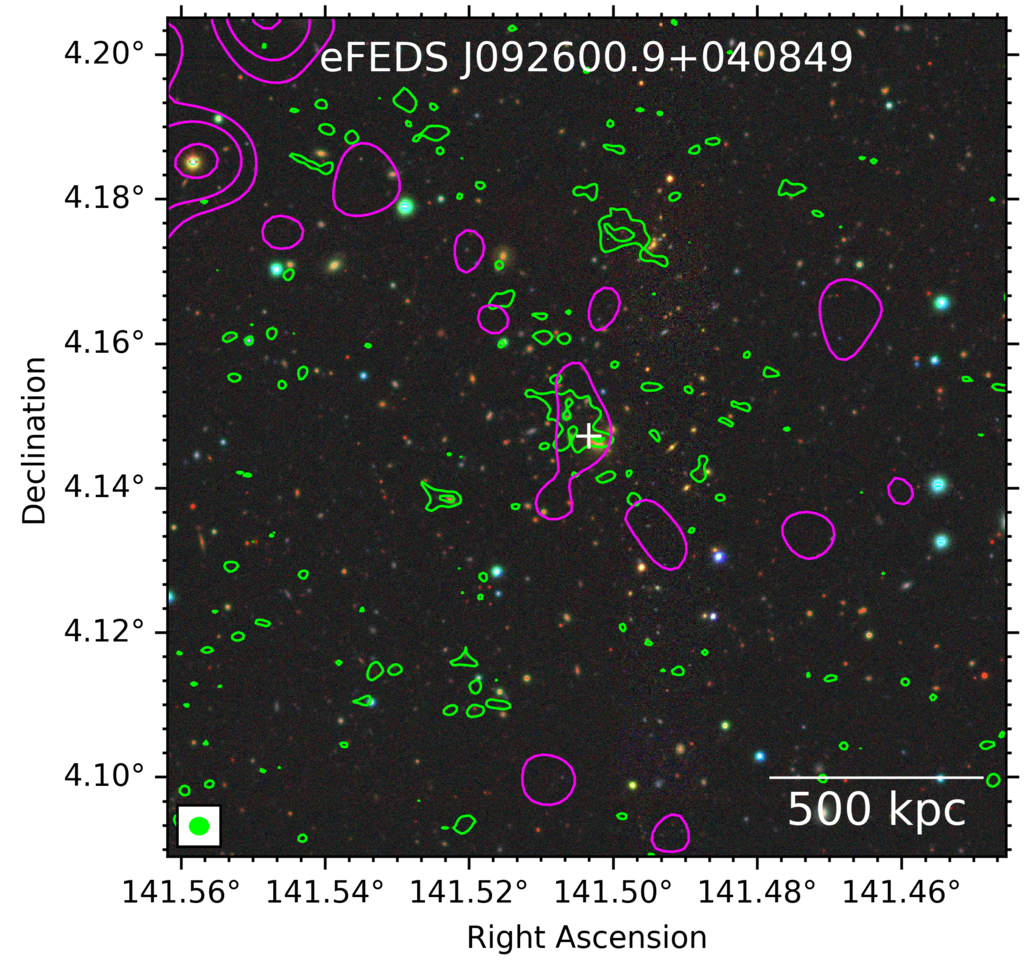}
  \caption{Clusters that host diffuse emission in the ICM. The background are DECaLS optical band-grz images. The LOFAR (green) and eROSITA (magenta) first contours are drawn at \textbf{$3\sigma$}. The subsequent contours are multiplied by a factor of 2. The cross at the center of the images indicates the location of the clusters. }
  \label{fig:radio_diffuse}
\end{figure*}

For completeness, we examined LOFAR radio and optical data of the Class~4 clusters to confirm that we did not miss any AGN-dominated BCGs in Class~4 clusters, where the counterpart finders {\sc NWAY} and MCMF agree on the reddest passive galaxy as the counterpart. We identified several interesting radio features in this class. In some of these clusters, we observe radio emission from an AGN, radio galaxy, and diffuse sources, such as halos and relics, associated with the cluster itself. These clusters are presented in Figures~\ref{fig:radio_diffuse} and \ref{fig:radioemission}. We describe the properties of the diffuse sources that could be radio halos or relics in the text below.

\begin{enumerate}
    \item {\bf eFEDS~J083826.4+015607 ($z=0.45$)}: The LOFAR observations detect a diffuse source with a projected size of $40$~arcsec, located  $2$\arcmin\ to the north of the cluster center (Figure \ref{fig:radio_diffuse}, left). The diffuse source does not have an optical counterpart in the DECaLS image. If the diffuse source is part of the cluster, its projected size is 270~kpc, and it can be classified as a candidate radio relic.

\vspace{2mm}
    \item{\bf eFEDS~J084149.8+014717 ($z=0.38$)}: A diffuse source is detected with LOFAR roughly $2.5$\arcmin\ to the north of the cluster center (Figure~\ref{fig:radio_diffuse}, middle). It has a projected size of $1.2\times0.5$\arcmin$^2$. Its major axis is elongated in the NE--SW direction and is roughly perpendicular to the line to the cluster center. The diffuse source is not entirely associated with any radio galaxies. It has a flux density of $21.3\pm2.2\,\text{mJy}$ and a radio power of $(12.3\pm1.3)\times10^{24}\,\text{W\,Hz}^{-1}$ ($k-$corrected) if it is located at the cluster redshift of $z=0.387$. The properties of the diffuse source are consistent with those of a radio relic. Future studies are required to confirm the nature of the source.
\vspace{2mm}
    \item {\bf eFEDS~J092600.8+040849 ($z=0.31$)}: Diffuse radio emission is detected in the central region of the cluster with LOFAR (Figure~\ref{fig:radio_diffuse}, right). The diffuse source with a projected size of $40\times20$~arcsec$^2$ is elongated in the NE--SW direction, which is roughly similar to the major axis of the X-ray emission seen with eROSITA. The flux density of the radio source is $5.0\pm0.7\,\text{mJy,}$ which corresponds to a radio power of $(1.78\pm0.25)\times10^{24}\,\text{W\,Hz}^{-1}$. The diffuse radio source is not associated with the BCG and could be the brightest part of a radio halo. This requires deeper observations for confirmation.
\end{enumerate}

\section{Results and discussion}
We presented the X-ray, optical, and radio properties of the clusters and groups of galaxies identified and classified in the eFEDS point source catalog presented in S21. These clusters or groups are misclassified as point-like sources by the \rosi source detection algorithm with an extent likelihood of zero. Our goal is to characterize this sample, to understand the selection of the eFEDS survey, and to find AGN-dominated clusters that are worth further follow-up investigations. The clusters in the eFEDS point source catalog were identified by running the red-sequence finder algorithm MCMF around the X-ray centroids. In total, S21 found 346 clusters and groups in the eFEDS point source sample, spanning a high-redshift range from 0.1 to 1.3. The overall distribution of the redshifts of the sample is similar to the extent-selected sample, but the median redshift of 0.40 is slightly higher than that of the extended sample. Additionally, 10 new high-z clusters at redshifts z~$>0.8$ are identified in this new sample. In addition to 433 clusters in the \citet{Liu2021} extent-selected catalog (taking into account the 20\% contamination), this brings the total number of clusters to 779 in the eFEDS field with a source density of 5.6~clusters per square degree. We note that the selection of this catalog and the extent-selected sample are quite different and should be treated with care. This catalog of clusters should not be combined with the extent-selected sample for studies involving the sample properties and the X-ray selection function provided in \citet{Liu2021}.

\begin{figure*}
\begin{center}
\includegraphics[width=0.49\textwidth]{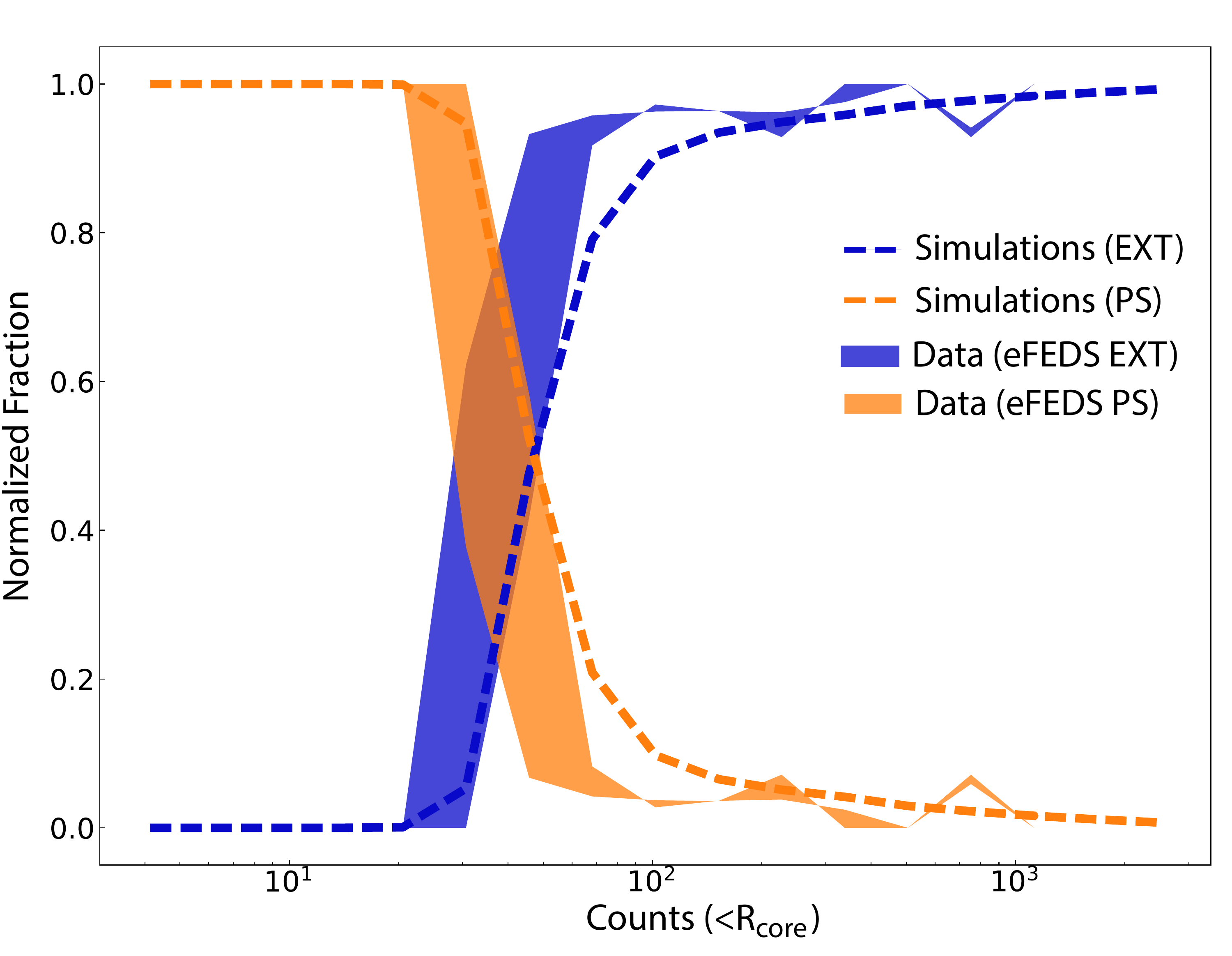}
\includegraphics[width=0.49\textwidth]{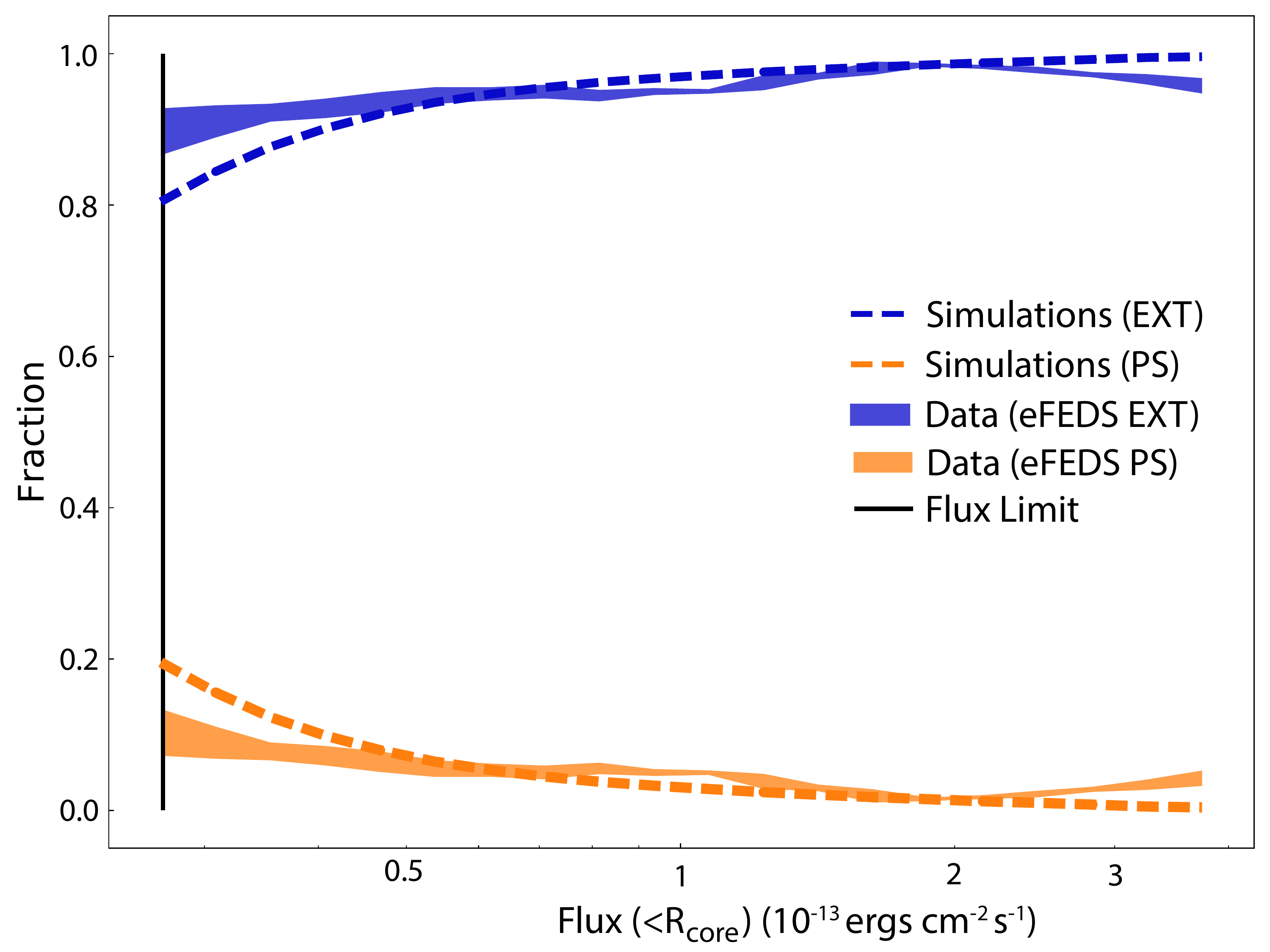}
\caption{Comparisons of the number of clusters in point source and extent selected catalogs with \rosi\ simulations. {\bf Left panel:} The fraction of clusters detected in the extent-selected sample vs. the clusters in the point source sample as a function of the counts in $R_{core}$ (specified as a ML\_CTS in the {\tt ermldet} detection runs) in simulations. {\bf Right panel:} The cumulative number of clusters in the extended and point source catalogs above a detection flux (ML\_Flux in the {\tt ermldet} runs) limit of 2.7$\times10^{-14}$ ergs~s$^{-1}$~cm$^{-2}$ measured within $R_{core}$ (EXT parameter in {\tt ermldet} outputs).}
\label{fig:sims}
\end{center}
\end{figure*}

Our first goal was to understand the reason for the misclassification of these clusters in the point source catalog by using the multiwavelength follow-up data in optical, X-ray, and radio bands. The reasons for this misclassification may include that this particular population of clusters is fainter, more compact, and hosts a cool-core or central AGN that dominates the overall X-ray emission compared to the extent-selected companions. The richness distribution of the clusters in the point source sample (after a similar optical cleaning was applied) implies that this sample may represent a similar population of clusters and groups as the extent-selected sample in terms of their optical properties. We were able to measure the luminosity within R$_{500}$ of 193 clusters with a detection confidence of $2\sigma$ or higher from the \rosi\ imaging analysis. Figure~\ref{fig:lumin} shows that 113 clusters lie below the flux limit of the eFEDS extent-selected clusters ($1.5\times10^{-14}$~ergs~s$^{-1}$~cm$^{-2}$). The luminosity and mass distribution of the sample shows that this sample is mostly dominated by low-mass groups and clusters that are underluminous and compact compared to the clusters in the extended source catalog. The comparison of gas density profiles out to R$_{500}$ (Figure~\ref{fig:density}) with the extent-selected clusters indicates that this sample is dominated by centrally bright clusters, either with an AGN at the core, or with bright cool cores. Comparing the distribution of their $R_{core}$ measurements and concentration parameters with the extent-selected sample, we can conclude that the majority of these clusters are indeed compact. The Kolmogorov-Smirnov test probability of the $R_{core}$ of the extent- and point-source-selected distributions to be drawn from the identical parent distribution is very low, $P\sim10^{15}$. Assuming the spectroscopic follow-up observations with SDSS (45\% of the total sources) are representative of the full sample, we find that AGNs (associated with clusters or in projection) constitute 10\% of the sample. In the few cases that AGNs or cool-cores dominate the ICM emission, it is possible that the overall flux is magnified by the central AGN and boosted above the detection limit. In these cases, the measured X-ray luminosities of the host clusters are biased high and should be treated as upper limits. Because of the PSF size of \rosi\ and the small extent of these clusters, we were not able to excise the core when we extracted X-ray properties. Although we find only a few cases like this, a follow-up study of these clusters with X-ray telescopes with better spatial resolution, for example, \xmm\ and \chandra, is required to distinguish the AGN emission from the cluster emission. Examining the correlation of their class, assigned based on the assigned counterparts, with luminosity and richness, we cannot pinpoint an isolated dominating effect, other than being compact and underluminous as the reason for the misclassification of these extended sources. Many factors seem to play a role in their selection.

To understand the selection and completeness of the extent-selected sample better, we compared the fraction of clusters detected as point sources with our simulations. These simulations were derived from the \rosi\ mock observations based on the MultiDark and UNIT N-body realizations of dark matter halos and the observed ICM models \citep{Comparat2020}. We ran our detection algorithm with the same set of settings applied to the data on the simulations at the full depth of the all-sky survey. The full depth was preferred because of its similarity to the depth of the eFEDS survey, but we note that the eFEDS survey is slightly deeper than the final \rosi\ all-sky survey at the same location. In Figure~\ref{fig:sims} we compare the fraction of sources detected in the extent-selected sample as a function of the detection counts (ML$\_$CTS in {\it ermdet} outputs) calculated within the extent parameter, which is related at the $R_{core}$ of the beta model \citep[see][for details]{Brunner2021}. Overall, the fraction of detected clusters in the observed eFEDS extent-selected sample (shaded regions) and simulations (dashed lines) agrees within the uncertainties. The clusters that are misclassified as point sources have a lower detection flux (or counts) than the extent-selected sample. The majority of the clusters (>50\%) are detected as point sources below a detection count limit of $\sim$40 or lower. This may mean that if a cluster has 40 counts or more in the core region, it is more likely to be detected as an extended source. If the cluster has 100 counts or more in the core, it has a $>90\%$ chance to be detected as an extended source. The rest, a small fraction of bright sources with 100 counts or more, may be detected as a point source if they are compact and at high redshift. The clusters below the flux limit may either have a bright cool core or AGN in their centers and/or have a very compact extended emission that is confused with a point source and is still detected by the boosted X-ray emission. The remaining clusters are fainter than the eFEDS extent-selected sample with a minimum flux of $\sim 2.7\times 10^{-14}$ ergs~s$^{-1}$~cm$^{-1}$ in the sample. These clusters are misclassified because of their lower fluxes, while their compactness, the central AGN, and their cool core might also play a minor role. The number of clusters below the detection flux limit of $2.7\times10^{-14}$ within $R_{core}$ is 56 in the point source sample. When we take the contamination of the 20\% level in the extended sample into account, the total number of genuine clusters increases to 489 (433+56) in the eFEDS field \citep{Liu2021}. From these numbers, we find that the extent selection finds roughly 88\% of the clusters. The simulations (see Figure~\ref{fig:sims} right-hand panel) predict this level to be $\sim$80\%, which agrees with our measurements at the 2$\sigma$ level. The small difference between the observations and simulations is expected and can be due to the spatially varying depth of the all-sky survey simulations used here. We recall that the eFEDS survey has a slightly higher and uniform depth. The scatter in the true versus observed flux may also cause this small departure from the expectations. 

We further examined this sample using the classification introduced by S21 and used our multi-wavelength observations. Based on the assigned optical counterparts by confirmation algorithms ({\sc NWAY} and MCMF), S21 classified the full sample into four categories. In Class~4, {\sc NWAY} and MCMF agree about the most probable optical counterpart, which is a red passive galaxy. On the other hand, in Class~3, {\sc NWAY} finds an AGN as the potential counterpart near the X-ray position. This class is more likely to host a bright AGN at the core of the host cluster.
We exploited a LOFAR 144 MHz observation of eFEDS to search for radio counterparts at the position of the BCG for the different classes of objects. We find that the radio detection fraction of Class~4 is higher ($\sim$47\%) than for other classes, as expected because this class represents secure clusters with an AGN at the position of the BCG. A second run of the algorithm was used to search for radio emission at the position of the AGN provided by {\sc NWAY}. The results for Class~4 are the same as in the first run, and the results for other classes are hard to interpret because of the relatively small statistics.

Another goal of this study was to identify clusters with AGN-dominated BCGs or cool-core clusters at high redshifts because X-ray surveys have often been criticized for missing high-redshift clusters and AGN-dominated groups in their extended source catalogs. We here established a new method for determining these sources in the point source catalog using their multiwavelength observations. 
To assess the cases in which extended sources were misclassified, we cross-matched the point source sample with the ACT and Planck SZ survey. Five clusters matched the ACT DR5 catalog at higher redshifts of $0.5-0.7,$ and one cluster matched the Planck PSZ2 catalog at a lower redshift (z=0.15). The measured photo-zs are consistent with the reported values in the SZ catalogs. The common properties of the matched clusters are that they all have a) a low signal-to-noise ratio SZ signal-to-noise ratio $<$6 and b) low luminosities and masses. This indicates that we probe high-z galaxy clusters and group populations with lower mass and lower luminosity with \rosi, close to the detection limit of ACT and Planck surveys. The majority of the clusters that we identified using our method are below the detection limit of the current public SZ surveys. Using this method, we would find not only massive high-redshift clusters, but also AGN-dominated compact galaxy groups at low redshift that are beyond the reach of the public data of the current SZ surveys. This demonstrates the power of \rosi\ over current public SZ surveys.

The method we established here to find groups and clusters that host bright AGN in their BCGs makes use of the cluster classification of S21. Starting from Class~3 objects, that is, the class of clusters or groups that have {\sc NWAY-} identified AGN at the same redshift as the host cluster, we downselected a sample of 46 clusters and groups that were brighter in the $g$ band than in the $r$ band, brighter in $z$ band than in $W1$ band, and brighter in $W1$ band than in $W2$ band  in their optical data than the rest of the cluster population. These clusters and groups are all listed in Table~\ref{tab:main} and are interesting targets in which to study AGN feedback in dense environments. We find that the radio detections of 33 of these clusters and groups are associated with the cluster itself within $R_500$ using the complementary LOFAR eFEDS data. The radio and infrared emission associated with these clusters indicates that the central supermassive black hole is indeed active and heats the surrounding medium to IR luminosities. Eight of these 33 clusters have radio emission in close proximity to the X-ray detection within $5^{\prime\prime}$ (source localization accuracy of eROSITA). The targeted by SDSS-V spectroscopic observations are available for five clusters and groups. These SDSS data provide the necessary high spectral resolution observations to characterize these sources in more detail and to test our classification method. The inspected spectra of these sources indeed show broad line signatures from the central AGN as well as the emission from the cluster galaxy, confirming that our selection method is successful in identifying these extraordinary clusters and groups. These populations of clusters and groups represent excellent targets for further astrophysical investigations and cosmology, for instance, studying the AGN feedback cycle in galaxy groups and clusters and constraining ultralight dark matter particles. One target we find through the search is a remarkable cluster and group, eFEDS~J091437.8+024558 (z=0.40, $M_{500}=3.61_{-0.36}^{+0.47}\times10^{14}$ $M_{sun}$), eFEDS~J083520.1+012516 (z=0.30, $M_{500}=1.15_{-0.18}^{+0.25}\times10^{14}$~$M_{sun}$), and  eFEDS~J092227.1+043339 (z=0.33, $M_{500}=6.55_{-1.63}^{+2.53}\times 10^{13}$ M$_{sun}$). This group of clusters is particularly bright in all the bands from X-rays and radio to infrared.  It is a BCG candidate, and could be an ideal target for studying AGN feedback. It might also be used to constrain the parameter space of axion-like particles.

\section{Conclusions}
Several studies (e.g., the Phoenix cluster) have shown that clusters of galaxies with compact cores and/or hosting a starburst galaxy and a strong point source at the center can be misclassified as point sources in the extent-selected X-ray surveys because their PSF is finite \citep{Green2017, McDonald2012, Somboonpanyakul2018}. As a pilot study, we examined the point sources in the benchmark \rosi\ Final Equatorial-Depth Survey (eFEDS) point source catalog. eFEDS is unique because the majority of the sources have spectroscopic data from the SDSS and GAMA surveys and deep radio data through LOFAR. We examined 346 clusters and groups identified via their red sequence by S21 through their multiwavelength observations in the optical, IR, radio, and X-ray bands in detail. We used these observations to understand the selection effects and the completeness of the eFEDS cluster sample. We find that the majority of the sources are underluminous and below the flux limit of the extended sources. Their faint fluxes and compact sizes cause them to be mischaracterized as point sources by our detection algorithm. Overall, the sample includes a small fraction ($\approx 10\%$) of clusters and groups hosting AGN at their centers, but this is not the main reason that these clusters are missed in the eFEDS extent-selected catalog. In the current version of the detection algorithm of {\tt eSASS}, a detected source with an extent likelihood parameter lower than 6 is set to have an extent likelihood of zero. Lowering the detection likelihood limit and allowing the extent likelihood parameter to be lower than 6 would help to classify these clusters as extended sources in future \rosi\ All Sky-survey catalogs. However, we note that this would also increase the contamination fraction in the extent-selected sample. 

Furthermore, by combining the cluster classification of S21, the optical spectra through SDSS and GAMA, the radio emission from LOFAR, infrared from the WISE survey, and X-ray data from \rosi, we developed a method for searching for groups and clusters with BCGs dominated by central AGN at their cores. This method was successfully tested in this work and resulted in the identification of eight low-mass groups and clusters that host AGN in their BCGs. Three of these are a promising low-mass cluster at $z=0.3-0.4$ for in-depth studies of AGN feedback. Most of these low-mass groups and clusters we identified are below the detection limit of the current SZ surveys and are undetected in their public catalogs. This highlights the power of \rosi\ in detecting this unique nearby low-mass, low-luminosity groups and high-redshift cool-core clusters, and verifies the validity of our method in identifying them in the point source catalog.

In the first \rosi\ All-Sky Survey (eRASS) in the German half of the sky, we expect to detect $\sim$10,000 extended sources and $\sim$500,000 point-like sources. With a simple scaling from eFEDS, we predict that $\sim$6,000 clusters and groups will be found in the point source catalog. Of these, 400 candidate clusters and groups will likely have an AGN in their BCGs. This sample will include not only high-redshift clusters, but also low-mass clusters and groups. \citet{Somboonpanyakul2021a} predicted a 2$\pm$1\% occurrence rate for the clusters that appear as point sources and are bright in mid-IR and radio wavelengths and 1\% occurrence rates for Phoenix-like clusters with a rapidly cooling core at high redshift. According to these rates, we expect to detect about $40-80$ high-redshift clusters of galaxies in this sample. A large sample with this size will allow studying cooling and feedback cycles and their evolution with redshift, discovering the new starburst-BCG clusters, and enabling searches for ultralight dark matter candidates, axion-like particles. However, identifying AGN in BCGs relies on the availability of large-area optical, radio, and IR surveys covering a significant fraction of the eRASS area. The large and uniform coverage with Legacy Survey DR10 in the g, r, z, and w bands, a dedicated follow-up program with SDSS (see Black Hole Mapper program \footnote{https://www.sdss5.org/mappers/black-hole-mapper}), and large coverage with radio surveys such as MeerKAT and ASKAP, eROSITA will provide a significant leap forward in the southern eRASS sky in identifying these clusters.

\begin{acknowledgement}
Authors thank the anonymous referee, Nicolas Clerc and Susanne Friedrich for helpful discussions and comments on the draft. 
\\
 E.B. acknowledges financial support from the European Research Council (ERC) Consolidator Grant under the European Union’s Horizon 2020 research and innovation programme (grant agreement CoG DarkQuest No 101002585). DNH acknowledges support from the ERC through the grant ERC-Stg DRANOEL n. 714245. JW acknowledges support by the Deutsche Forschungsgemeinschaft (DFG, German Research Foundation) under Germany’s Excellence Strategy - EXC-2094 -390783311.
 
\\

This work is based on data from eROSITA, the soft X- ray instrument aboard SRG, a joint Russian-German science mission supported by the Russian Space Agency (Roskosmos), in the interests of the Russian Academy of Sciences represented by its Space Research Institute (IKI), and the Deutsches Zentrum f{\"{u}}r Luft und Raumfahrt (DLR). The SRG spacecraft was built by Lavochkin Association (NPOL) and its subcontractors, and is operated by NPOL with support from the Max Planck Institute for Extraterrestrial Physics (MPE).

The development and construction of the eROSITA X-ray instrument was led by MPE, with contributions from the Dr. Karl Remeis Observatory Bamberg \& ECAP (FAU Erlangen-Nuernberg), the University of Hamburg Observatory, the Leibniz Institute for Astrophysics Potsdam (AIP), and the Institute for Astronomy and Astrophysics of the University of T{\"{u}}bingen, with the support of DLR and the Max Planck Society. The Argelander Institute for Astronomy of the University of Bonn and the Ludwig Maximilians Universit{\"{a}}t Munich also participated in the science preparation for eROSITA.

The eROSITA data shown here were processed using the {\tt eSASS} software system developed by the German eROSITA consortium.
\\

Funding for the Sloan Digital Sky Survey IV has been provided by the Alfred P. Sloan Foundation, the U.S. Department of Energy Office of Science, and the Participating Institutions. 
SDSS-IV acknowledges support and resources from the Center for High Performance Computing  at the University of Utah. The SDSS website is www.sdss.org. SDSS-IV is managed by the Astrophysical Research Consortium for the Participating Institutions of the SDSS Collaboration including  the Brazilian Participation Group, the Carnegie Institution for Science,  Carnegie Mellon University, Center for Astrophysics | Harvard \& Smithsonian, the Chilean Participation Group, the French Participation Group, Instituto de Astrof\'isica de Canarias, The Johns Hopkins University, Kavli Institute for the Physics and Mathematics of the Universe (IPMU) / University of Tokyo, the Korean Participation Group, 
Lawrence Berkeley National Laboratory, Leibniz Institut f\"ur Astrophysik Potsdam (AIP),  Max-Planck-Institut f\"ur Astronomie (MPIA Heidelberg), Max-Planck-Institut f\"ur Astrophysik (MPA Garching), Max-Planck-Institut f\"ur Extraterrestrische Physik (MPE), National Astronomical Observatories of China, New Mexico State University, New York University, University of Notre Dame, Observat\'ario Nacional / MCTI, The Ohio State 
University, Pennsylvania State University, Shanghai Astronomical Observatory, United Kingdom Participation Group, Universidad Nacional Aut\'onoma de M\'exico, University of Arizona, University of Colorado Boulder, University of Oxford, University of Portsmouth, University of Utah, University of Virginia, University of Washington, University of 
Wisconsin, Vanderbilt University, and Yale University.

\\

This work made use of SciPy \citep{jones_scipy_2001}, matplotlib, a Python library for publication quality graphics \citep{Hunter:2007}, Astropy, a community-developed core Python package for Astronomy \citep{2013A&A...558A..33A}, NumPy \citep{van2011numpy}. 
\end{acknowledgement}

\bibliography{literature.bib}

\begin{appendix}
\onecolumn

\section{Multiwavelength images of clusters with radio emission}

\begin{figure*}[ht]
\centering
 \captionsetup[subfigure]{labelformat=empty}
     \centering
     \subfloat[][]{\includegraphics[width=0.9\columnwidth]{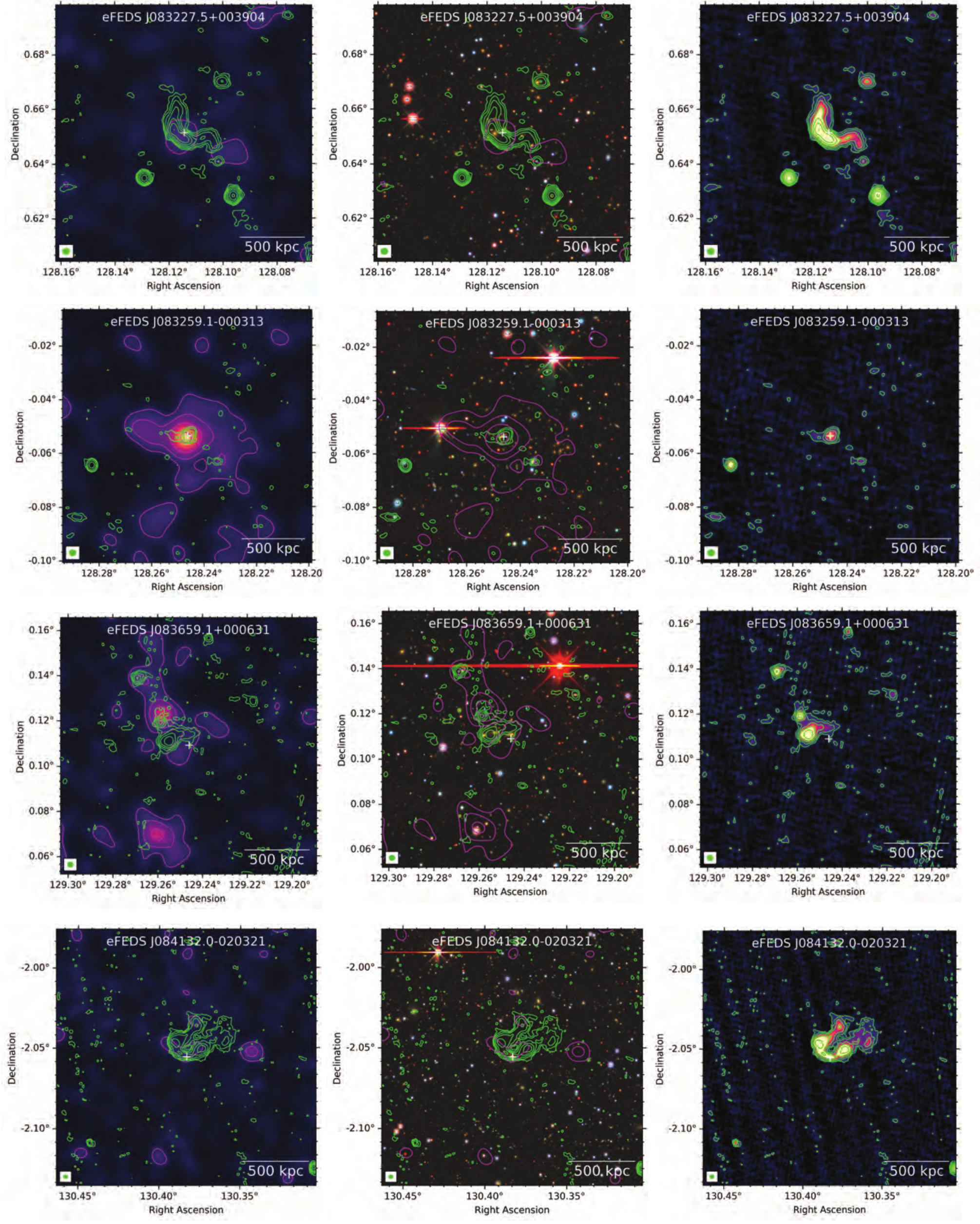}}
    
\caption{\label{fig:radioemission}{\it Left panel}: Soft-band $0.5-2$~keV \rosi\ images of the clusters that shown significant radio emission associated with the BCG. The overlaid green contours mark the radio emission from LOFAR data. Middle panel: DECALS images of the same clusters. Right panel: LOFAR images. }
 \end{figure*}
\begin{figure*}[t]
     \ContinuedFloat
     \captionsetup[subfigure]{labelformat=empty}
     \centering
  
    \subfloat[][]{\includegraphics[width=1\columnwidth]{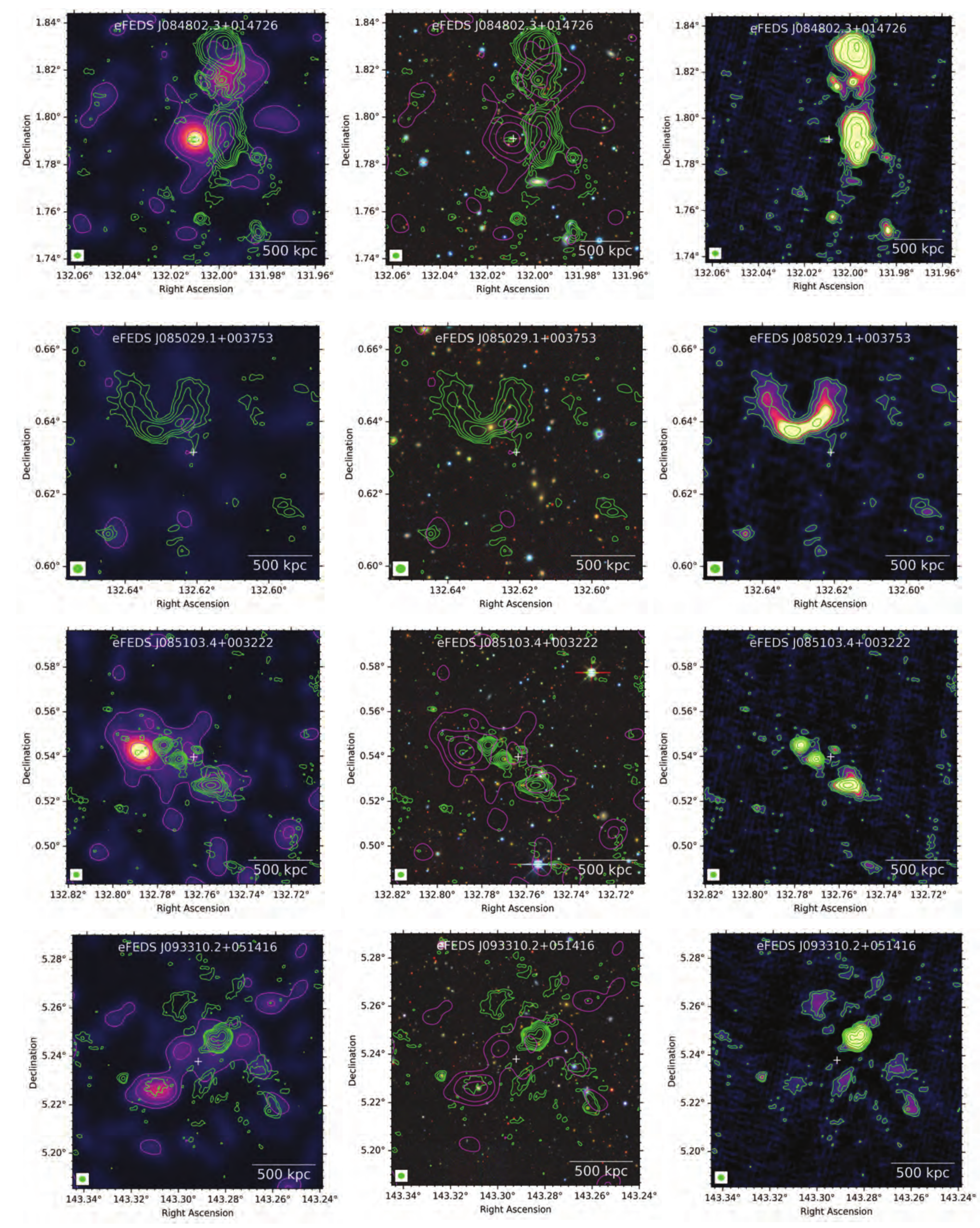}}
     \caption{Continued.}
 \end{figure*}
\begin{figure*}[t]
     \ContinuedFloat
     \captionsetup[subfigure]{labelformat=empty}
     \centering

       \subfloat[][]{\includegraphics[width=1\columnwidth]{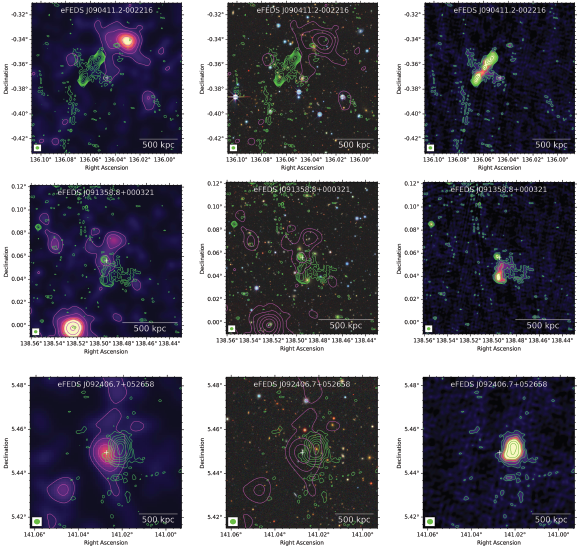}}
     \caption{Continued.}
 \end{figure*}

\newpage
\onecolumn
\fontsize{8.0pt}{0cm}\selectfont
\begin{landscape}
\begin{longtable}{lcccccccccccp{0.25\linewidth}}
\caption{\label{tab:main} X-ray, Optical, and Radio properties of the clusters in the eFEDS point source catalog.} \\
\hline\hline
ID  & RA & Dec & $z$  & Rich. & Class & R$_{500}$  & L$_{500}$ & M$_{500}$ & $F_{\rm radio, tot}$ & Chandra & XMM & Notes \\

(eFEDS~+) & [deg] & [deg] & & & & [kpc] & [$10^{42}$~ergs~s$^{-1}$] & [$10^{13}$~$M_{sun}$] & [$10^{-2}$~Jy] & & & \\
\hline
\endfirsthead
\caption{continued.}\\
\hline\hline
ID  & RA & Dec & $z$  & Rich. & Class & R$_{500}$  & L$_{500}$ & M$_{500}$ & $F_{\rm radio, tot}$ & Chandra & XMM & Notes \\
\hline
(eFEDS~+) & [deg] & [deg] & & & & [kpc] & [$10^{42}$~ergs~s$^{-1}$] & [$10^{13}$~$M_{sun}$] & [$10^{-2}$~Jy] & & & \\
 \hline
\endhead
\hline
\endfoot
J082821.3-000239 &  $ 127.0889 $ &  $-0.0444 $ & 0.71 & 55 & 5 &  $ 465_{-138}^{+95} $ &  $ 9.32_{-7.36}^{+10.78} $ &  $ 6.31_{-4.11}^{+4.69} $ &  &  N &  N &   \\
J082956.6-000223 &  $ 127.4858 $ &  $-0.0400 $ & 0.44 & 28 & 3 &  $ 346_{-177}^{+122} $ &  $ 1.07_{-0.99}^{+2.35} $ &  $ 1.88_{-1.66}^{+2.77} $ &  $ 0.24\pm 0.08 $ &  N &  N &   \\
J083018.9+015456 &  $ 127.5788 $ &  $+1.9156 $ & 0.39 & 37 & 2 &  $ 607_{-101}^{+89} $ &  $ 12.50_{-6.26}^{+9.40} $ &  $ 9.60_{-4.04}^{+4.80} $ &  $ 6.12\pm 0.14 $ &  N &  N &   \\
J083026.9+010115 &  $ 127.6123 $ &  $+1.0209 $ & 1.32 & 175 & 3 &  $ 524_{-41}^{+50} $ &  $ 98.80_{-34.00}^{+47.20} $ &  $ 18.20_{-3.90}^{+5.80} $ &  $ 0.46\pm 0.05 $ &  N &  N &   \\
J083056.4-020032$^{\dagger}$ &  $ 127.7352 $ &  $-2.0090 $ & 0.30 & 18 & 3 &  $ 575_{-70}^{+59} $ &  $ 8.47_{-3.60}^{+3.83} $ &  $ 7.39_{-2.38}^{+2.52} $ &  $ 0.52\pm 0.09 $ &  N &  N &   \\
J083101.7+020007 &  $ 127.7574 $ &  $+2.0021 $ & 0.42 & 25 & 5 &  $ 279_{-166}^{+160} $ &  $ 0.45_{-0.44}^{+2.35} $ &  $ 0.98_{-0.91}^{+2.81} $ &  &  N &  N &   \\
J083104.1-015741$^{\dagger}$ &  $ 127.7675 $ &  $-1.9615 $ & 0.28 & 15 & 3 &  $ 605_{-45}^{+56} $ &  $ 9.71_{-2.53}^{+4.09} $ &  $ 8.42_{-1.77}^{+2.58} $ &  $ 1.95\pm 0.08 $ &  N &  N &   \\
J083134.3+015601 &  $ 127.8932 $ &  $+1.9337 $ & 1.21 & 126 & 2 &  $ 560_{-117}^{+109} $ &  $ 19.10_{-12.20}^{+19.80} $ &  $ 10.40_{-5.26}^{+7.30} $ &  &  N &  N &   \\
\hline
\end{longtable}
\tablefoot{In columns 11 and 12, we note whether the cluster has been observed by {\sl Chandra} and {\sl XMM-Newton} as of December, 2021. IDs marked with $\dagger$ are the clusters which have counterparts that show significant color offsets from the rest of the population with the selection criteria of $g - r <1.15$, and $W1 - W2>-0.3$.}
\end{landscape}

\end{appendix}

\end{document}